\documentclass[aps,prd,onecolumn,eqsecnum, nofootinbib]{revtex4-2}
\usepackage{bm}
\usepackage{amssymb}
\usepackage{amsmath}
\usepackage{amsfonts}
\usepackage{graphicx}
\usepackage{hyperref}
\DeclareSymbolFont{EulerScript}{U}{eus}{m}{n}
\SetSymbolFont{EulerScript}{bold}{U}{eus}{b}{n}
\DeclareSymbolFontAlphabet\scrpt{EulerScript}
\newcommand{\m}{{\sf m}}
\newcommand{\elltide}{{\mbox{\scriptsize $\ell$-tide}}} 
\newcommand{\ellmass}{{\mbox{\scriptsize $\ell$-mass}}} 
 
\newcommand{\A}{{\scrpt A}} 
 
\newcommand{\LL}{{\scrpt L}}

\newcommand{\gotha}{\mathfrak{A}} 
\newcommand{\gothb}{\mathfrak{B}} 
\newcommand{\gothc}{\mathfrak{C}} 
\newcommand{\gothd}{\mathfrak{D}} 
\newcommand{\rr}{\mathfrak{r}} 
\newcommand{\OO}{\mathfrak{O}} 
\newcommand{\N}{\mathfrak{N}}
\newcommand{\stf}[1]{{\langle #1 \rangle}} 
\newcommand{\e}[1]{{$\times 10^{#1}$}}
\allowdisplaybreaks
\begin{document}
\title{General relativistic dynamical tides in binary inspirals, without modes} 
\author{Tristan Pitre and Eric Poisson}  
\affiliation{Department of Physics, University of Guelph, Guelph, Ontario, N1G 2W1, Canada} 
\date{January 30, 2024} 
\begin{abstract}
A neutron star in an inspiraling binary system is tidally deformed by its companion, and the effect leaves a measurable imprint on the emitted gravitational waves. While the tidal interaction falls within the regime of static tides during the early stages of inspiral, a regime of dynamical tides takes over in the later stages. The description of dynamical tides found in the literature makes integral use of a spectral representation of the tidal deformation, in which it is expressed as a sum over the star's normal modes of vibration. This description is deeply rooted in Newtonian fluid mechanics and gravitation, and we point out that considerable obstacles manifest themselves in an extension to general relativity. To remedy this we propose an alternative, mode-less description of dynamical tides that can be formulated in both Newtonian and relativistic mechanics. Our description is based on a time-derivative expansion of the tidal dynamics, in which the external, orbital time scale associated with the tidal field is taken to be long compared with the internal, hydrodynamical time scale associated with the neutron star. The tidal deformation is characterized by two sets of Love numbers: the familiar static Love numbers $k_\ell$, which appear in the regime of static tides, and the dynamic Love numbers $\ddot{k}_\ell$, which emerge in the regime of dynamical tides. These numbers are computed here for polytropic stellar models in both Newtonian gravity and general relativity. On the face of it, the time-derivative expansion of the tidal dynamics seems to preclude any attempt to capture an approach to resonance, which occurs when the frequency of the tidal field becomes equal to a normal-mode frequency; such an approach is the main reason for which the regime of dynamical tides becomes important in the late inspiral, and it is fully captured by the mode representation of the tidal deformation. To overcome this limitation we propose a pragmatic extension of the time-derivative expansion which does capture an approach to resonance. We demonstrate that with this extension, our formulation of dynamical tides should be just as accurate as the $f$-mode truncation of the mode representation, in which the sum over modes is truncated to a single term involving the star's fundamental mode of vibration.       
\end{abstract} 
\maketitle

\section{Introduction and summary} 
\label{sec:intro} 

\subsection{Tidal deformations in binary inspirals} 

The tidal deformation of a neutron star, as it occurs in the late stages of a binary inspiral driven by gravitational radiation reaction, makes an imprint on the emitted gravitational waves. It was recognized long ago \cite{flanagan-hinderer:08} that a measurement of this effect could constrain the equation of state of nuclear matter at high densities, which is poorly understood \cite{ozel-freire:16, oertel-etal:17, baym-etal:18}. A measurement of the tidal deformability of a neutron star was attempted in the case of GW170817 \cite{GW170817:17, GW170817:18, narikawa-uchikata-tanaka:21}, and the resulting upper bound favors a soft equation of state that produces a relatively small neutron star \cite{landry-essick-reed-chatziioannou:20}. A detailed review of these developments was crafted by Chatziioannou \cite{chatziioannou:20}, and prospects for future measurements of this sort are summarized in Ref.~\cite{pacilio-maselli-fasano-pani:22}.

The aspects of the tidal interaction between a neutron star and its companion that are most readily accessible to gravitational-wave measurements fall within the regime of {\it static tides} (or adiabatic tides), in which time derivatives of the tidal field can be neglected. However, it was pointed out \cite{hinderer-etal:16, steinhoff-etal:16} that the regime of {\it dynamical tides}, in which the time derivatives are not neglected, will soon be accessible to measurement, and will yield additional insights into the internal constitution of neutron stars. This observation has prompted a vigorous investigation of this regime, going from detailed models of a dynamically deformed neutron star \cite{steinhoff-etal:21, andersson-pnigouras:21, passamonti-andersson-pnigouras:21, passamonti-andersson-pnigouras:22}, to models of gravitational waveforms \cite{schmidt-hinderer:19, mandal-etal:23, mandal-etal:24}, to prospects for future detections and what can be learned from them \cite{andersson-ho:18, williams-pratten-schmidt:22, pratten-schmidt-williams:22}. The regime of dynamical tides was also extended to include the gravitomagnetic sector of the tidal dynamics \cite{flanagan-racine:07, poisson:20a, gupta-steinhoff-hinderer:21, gupta-steinhoff-hinderer:23}, and it was included in a description of the $p$-$g$-mode instability of neutron stars \cite{weinberg:16}. 

All the work on dynamical tides reviewed in the preceding paragraph relies on a spectral representation of the tidal deformation, in which it is expressed as a sum over the normal modes of vibration of a neutron star. Our purpose with this paper is (i) to make the point that such a description is deeply rooted in Newtonian fluid mechanics and gravitation and that considerable obstacles manifest themselves in a generalization to relativistic mechanics, and (ii) to present an alternative, mode-less description that can be ported to general relativity. Our approach, we argue, provides a practical and reliable way to model dynamical tides in general relativity. We elaborate on these points, and summarize our results, in the remainder of this introductory section. 

\subsection{Mode description: Frequency-domain Love numbers}

To begin our discussion, we consider a tidally deformed neutron star of mass $M$ and radius $R$ within the framework of Newtonian fluid mechanics and gravitation; the star is assumed to be nonrotating. The tidal  deformation is described by a Lagrangian displacement vector $\bm{\xi}(t, \bm{x})$, which takes a fluid element at its original position $\bm{x}$ and places it at the perturbed position $\bm{x} + \bm{\xi}$. The displacement vector is decomposed in terms of the star's normal modes of vibration (details can be found, for example, in Sec.~2.5.3 of Ref.~\cite{poisson-will:14}), as
\begin{equation}
\bm{\xi}(t, \bm{x}) = \sum_{K} q_K(t)\, \bm{\zeta}_K(\bm{x}),
\label{xi_mode} 
\end{equation}
in which $q_K(t)$ is the mode amplitude and $\bm{\zeta}_K(\bm{x})$ is the mode eigenfunction; $K$ is a mode label, and the (formally infinite) sum extends over all modes. The tidal acceleration $\bm{g} = \bm{\nabla} U^{\rm tidal}$ is expanded in a similar way ($U^{\rm tidal}$ is the gravitational potential created by the companion), and the fluid equations imply that each mode behaves as a driven harmonic oscillator, with an equation of motion
\begin{equation}
\ddot{q}_K + \omega_K^2 q_K = g_K,
\end{equation}
where overdots indicate differentiation with respect to time, and $\omega_K$ is the mode frequency. The solution is readily expressed in the frequency domain, and we have that
\begin{equation}
\tilde{q}_K(\omega) = \omega_K^{-2}\, \A_K\, \tilde{g}_K(\omega), \qquad
\A_K := \bigl( 1 - \omega^2/\omega_K^2 \bigr)^{-1}.  
\end{equation}
This solution features the familiar response function of an oscillator, which diverges when the frequency $\omega$ of the tidal force becomes equal to a mode frequency $\omega_K$. The importance of the dynamical aspects of the tidal deformation takes its origin in the amplification factor $\A_K$. In the context of a binary inspiral, with a typical tidal frequency of 500 Hz and a typical $f$-mode frequency of 1000 Hz, we have that $\A \simeq 1.19$, giving rise to a 20\% enhancement of the tidal deformation.

The displacement vector is used to compute the multipole moments of the deformed mass distribution. To describe this we refine our notation and let $n\ell\m$ stand for the mode label $K$. Here, $\ell$ is the multipolar order (starting at $\ell=2$ with the quadrupole moment), $\m$ is the azimuthal index (ranging from $-\ell$ to $+\ell$), and $n$ labels the mode overtones for each multipolar order (starting at $n=0$ with the $f$-mode, ranging over positive integers for the $p$-modes, and over negative integers for the $g$-modes). For a multipole of order $\ell$ we write the frequency-domain tidal potential as [Eq.~(\ref{tidal_potential2})] 
\begin{equation}
\tilde{U}^{\rm tidal}(\omega,\bm{x}) = -\frac{1}{(\ell-1)\ell} \tilde{\cal E}^{\ell\m}(\omega)\, r^\ell Y^{\ell\m}(\theta,\phi),
\label{tidal_potential} 
\end{equation}
where $\tilde{\cal E}^{\ell\m}(\omega)$ are the (Fourier transform of the) tidal multipole moments, $r$ is the distance to the star's center-of-mass, and $Y^{\ell\m}(\theta,\phi)$ are spherical harmonics. The star's mass multipole moments are then given by [Eq.~(\ref{k_omega_def})] 
\begin{equation}
\tilde{\cal Q}^{\ell\m}(\omega) = -\frac{2(\ell-2)!}{(2\ell-1)!!} G^{-1} R^{2\ell+1}\, \tilde{k}_\ell(\omega)\,
\tilde{\cal E}^{\ell\m}(\omega),
\label{Q_vs_E_freq} 
\end{equation} 
where $G$ is the gravitational constant, $R$ is the star's radius, and $\tilde{k}_\ell(\omega)$ is a frequency-domain Love number given by [Eq.~(\ref{k_modesum})]
\begin{equation}
\tilde{k}_\ell(\omega) = \frac{2\pi \ell^2}{2\ell+1} 
\sum_n \frac{GM/R^3}{\omega_{n\ell}^2}\, \A_{n\ell}\, \frac{\OO_{n\ell}^2}{\N_{n\ell}}, 
\label{k_mode} 
\end{equation}
where $\omega_{n\ell}$ is the frequency of the $n\ell\m$ mode, $\A_{n\ell} := (1 - \omega^2/\omega_{n\ell}^2)^{-1}$ is the amplification factor for this mode, $\OO_{n\ell}$ is an overlap integral between the tidal acceleration $\bm{g}$ and the mode eigenfunction $\bm{\zeta}_{n\ell}$ --- refer to  Eq.~(\ref{O_nell}) for a precise definition --- and $\N_{n\ell}$ is a normalization factor --- see Eq.~(\ref{N_nell2}). All quantities that appear in Eq.~(\ref{k_mode}), including the combination $\omega_{n\ell}^2/(GM/R^3)$, are dimensionless. We see once more the impact of the amplification factor: the Love number increases significantly when $\omega$ becomes comparable to a mode frequency $\omega_{n\ell}$.

\subsection{Key ingredients of the Newtonian mode description, and obstacles to a relativistic generalization}
\label{subsec:obstacles} 

The mode description of a star's tidal deformation relies on three key ingredients. As we shall explain, these are deeply rooted in the Newtonian framework (fluid mechanics and gravitation), and they are not readily ported to general relativity. Until these obstacles are overcome, a mode description of tidal deformation must remain foreign to relativistic mechanics and gravitation, and a suitable alternative must be identified.

The first ingredient is a precise and unambiguous identification of the tidal acceleration $\bm{g}$, which is used to compute the mode projections $g_K$. In the Newtonian framework, the identification relies on a unique partition of $U$, the complete gravitational potential, into a piece $U^{\rm tidal}$ created entirely by the star's companion, and another piece $U^{\rm star}$ created by the star. Such a partition is possible because the governing equation is linear; this is Poisson's equation $\nabla^2 U = -4\pi G \rho$, where $\rho$ is the mass density. The tidal potential is a solution to Laplace's equation, $\nabla^2 U = 0$, in a region of space that includes the star's interior and a portion of its exterior --- it is cut off to exclude the companion. In this region the solution is given globally by Eq.~(\ref{tidal_potential}) after performing a multipole expansion. The tidal acceleration is then $\bm{g} = \bm{\nabla} U^{\rm tidal}$. 

Such a partition is not available in general relativity, because of the nonlinear nature of the Einstein field equations. The metric tensor $g_{\alpha\beta}$ cannot, in general, be decomposed into a piece created entirely by the star, and a remaining piece created by its companion; both pieces are intricately linked by the theory's nonlinearities. Now, the metric outside a tidally deformed body can still be computed and presented as a multipole expansion (as was done, for example, in Ref.~\cite{poisson:21a}), but as Sam Gralla pointed out \cite{gralla:18}, it cannot be partitioned uniquely into star and tidal pieces; the partition is necessarily ambiguous. Even if a ``preferred'' partition could be identified, it is not at all clear how the tidal piece of the metric could be extended from the stellar exterior to its interior so as to provide an analogy with $U^{\rm tidal}$. It is not clear, in particular, that such an extension would produce a nonsingular metric throughout the stellar interior. Thus, a first obstacle to the formulation of a relativistic spectral representation of the tidal deformation is the absence of a clear path to define a purely tidal field within the stellar interior.

A second key ingredient implicated in a mode description of the tidal deformation is the existence of an inner product for mode eigenfunctions. The inner product is invoked when defining the mode projection $g_K$ of the tidal acceleration, which is given schematically by 
\begin{equation}
g_K = \int \rho\, \bm{g} \cdot \bm{\zeta}_K\, dV;
\label{g_mode}
\end{equation}
the precise definition is given by Eq.~(\ref{gK}) below. The main point is that in the Newtonian framework, the inner product is defined by an integral over a bounded region of space, the volume occupied by the star. 

The situation is very different in general relativity. In this setting, the dynamical degrees of freedom associated with the tidal deformation include fluid variables, which are defined entirely within the star, but also gravitational-field variables, which are defined everywhere. The symplectic form in the phase space of these degrees of freedom (see Ref.~\cite{friedman:78}, or Sec.~7.4 of Ref.~\cite{friedman-stergioulas:13}) could be exploited to define a notion of inner product, but this will necessarily include contributions from the field variables, given by integrals over all space. In this context, it is not clear how an equation such as Eq.~(\ref{g_mode}) would generalize, and whether it would require a ``tidal metric'' that is defined everywhere in spacetime. In view of this, it is not clear how a spectral representation of tidal deformation could be formulated in general relativity, and whether it could ever be turned into a practical method of computation. A second obstacle to a general relativistic formulation is therefore the complicated nature of the inner product. 

A third key ingredient is the fact that in the Newtonian theory, the normal modes of vibration form a complete set of basis functions to represent {\it any} conceivable perturbation \cite{beyer-schmidt:95}. This ensures that nothing can be missed when the perturbation is expressed as in Eq.~(\ref{xi_mode}). The situation is very different in general relativity: the normal modes become quasi-normal modes, and they are known to be incomplete \cite{kokkotas-schmidt:99}. This implies that some perturbations {\it cannot} be expressed as a sum over modes, and that an expansion such as Eq.~(\ref{xi_mode}) may indeed miss something. In practice this limitation may not be too serious, because we are interested in a very specific type of perturbation, a tidal deformation, and it could be that in this case, a representation in terms of modes would prove to be perfectly adequate. It is, nevertheless, a cause for concern that the spectral representation may not be sufficiently general, and that it may be difficult to judge in practice whether it is sufficiently accurate in a given situation. Thus, a third obstacle toward a mode description of tidal deformation in general relativity is the formal absence of mode completeness.

Our conclusion is that while the spectral representation is a conceptually powerful and practical method to describe a tidal deformation in Newtonian mechanics, it does not readily generalize to a relativistic setting. At the very least, some challenging obstacles must be overcome before a solid foundation is secured, and it could well be that any resulting formalism will have to rely on approximations. Given this state of affairs, it appears to us crucial to offer an alternative description of dynamical tides, without modes. 

\subsection{Mode-less description of dynamical tides: Newtonian theory} 

Our alternative description of dynamical tides is based on Eq.~(\ref{Q_vs_E_freq}), in which we expand $\tilde{k}_\ell(\omega)$ in powers of $\omega^2$, and which then becomes 
\begin{equation}
{\cal Q}^{\ell\m}(t) = -\frac{2(\ell-2)!}{(2\ell-1)!!} G^{-1} R^{2\ell+1}
\biggl[ k_\ell\, {\cal E}^{\ell\m}(t) - \ddot{k}_\ell\, \frac{R^3}{GM}\, \ddot{\cal E}^{\ell\m}(t) + \cdots \biggr] 
\label{Q_vs_E_time} 
\end{equation} 
after inverting the Fourier transform [Eq.~(\ref{Q_vs_E_dyn})]. Overdots on ${\cal E}^{\ell\m}(t)$ indicate differentiation with respect to time, the ellipsis represents omitted higher-derivative terms, and we have introduced [Eq.~(\ref{k_ddotk})] 
\begin{equation}
k_\ell := \tilde{k}_\ell(\omega = 0), \qquad
\ddot{k}_\ell := \frac{GM}{R^3} \frac{d \tilde{k}_\ell}{d\omega^2} \biggr|_{\omega = 0}
\end{equation}
as {\it static} and {\it dynamic} Love numbers, respectively (the overdots on $k_\ell$ do not indicate differentiation with respect to time). Equation (\ref{Q_vs_E_time}) corresponds to a low-frequency approximation of Eq.~(\ref{Q_vs_E_freq}); the approximation is valid when the external frequency $\omega$ is low compared with any mode frequency $\omega_{n\ell}$. Alternatively, and more germane to our mode-less point of view, Eq.~(\ref{Q_vs_E_time}) is an expansion in powers of
\begin{equation} 
\epsilon := \frac{\mbox{internal, hydrodynamical time scale}}{\mbox{external, orbital time scale}} \ll 1, 
\end{equation} 
with an internal time scale comparable to $(R^3/GM)^{1/2}$, and an external time scale comparable to $(d^3/GM_{\rm tot})^{1/2}$, where $d$ is the typical distance to the companion, and $M_{\rm tot}$ is the sum of masses (neutron star and companion); for comparable masses we have that $\epsilon$ is small whenever $R/d$ is small, which is realized during most of the inspiral. The link between the time-derivative and low-frequency expansions is provided by the fact that $\omega$ is of the order of $(GM_{\rm tot}/d^3)^{1/2}$, the reciprocal of the external time scale, while $\omega_{n\ell}$ is of the order of\footnote{The numerical factor between $\omega_{n\ell}$ and $(GM/R^3)^{1/2}$ varies substantially as we move along the sequence of modes, with $p$-modes increasing in frequency, and $g$-modes decreasing in frequency.} $(GM/R^3)^{1/2}$, the reciprocal of the internal time scale; it then follows that $\epsilon \sim \omega/\omega_{n\ell}$.

\begin{table}
\caption{\label{tab:love}Static and dynamic Love numbers for stellar models with an equation of state $p = K \rho^{1+1/n}$.} 
\begin{ruledtabular}
\begin{tabular}{cccc}
  $n$ & $\ell$ & $k_\ell$ & $\ddot{k}_\ell$ \\
  \hline
1.0 & 2 & 2.599089$\times 10^{-1}$ & 1.726055$\times 10^{-1}$ \\
      & 3 & 1.064540$\times 10^{-1}$ & 3.688452$\times 10^{-2}$ \\
      & 4 & 6.024126$\times 10^{-2}$ & 1.450994$\times 10^{-2}$ \\
      & 5 & 3.929250$\times 10^{-2}$ & 7.352852$\times 10^{-3}$ \\
  & & & \\
1.5 & 2 & 1.432776$\times 10^{-1}$ & 6.745865$\times 10^{-2}$ \\
      & 3 & 5.284790$\times 10^{-2}$ & 1.405281$\times 10^{-2}$ \\
      & 4 & 2.739262$\times 10^{-2}$ & 5.334982$\times 10^{-3}$ \\
      & 5 & 1.656842$\times 10^{-2}$ & 2.599620$\times 10^{-3}$ \\
  & & & \\
2.0 & 2 & 7.393839$\times 10^{-2}$ & 2.416719$\times 10^{-2}$ \\
      & 3 & 2.439400$\times 10^{-2}$ & 4.972947$\times 10^{-3}$ \\
      & 4 & 1.150775$\times 10^{-2}$ & 1.824720$\times 10^{-3}$ \\
      & 5 & 6.419967$\times 10^{-3}$ & 8.530139$\times 10^{-4}$ \\
  & & & \\
2.5 & 2 & 3.485234$\times 10^{-2}$ & 7.745541$\times 10^{-3}$ \\
      & 3 & 1.019200$\times 10^{-2}$ & 1.595238$\times 10^{-3}$ \\
      & 4 & 4.341510$\times 10^{-3}$ & 5.647037$\times 10^{-4}$ \\
      & 5 & 2.220149$\times 10^{-3}$ & 2.520272$\times 10^{-4}$
\end{tabular}  
\end{ruledtabular} 
\end{table}

In our derivation of Eq.~(\ref{Q_vs_E_time}), the absence of a term proportional to $\dot{\cal E}^{\ell\m}(t)$ came as a consequence of the fact that $\tilde{k}_\ell(\omega)$ is actually a function of $\omega^2$; this can be gathered from the mode representation of Eq.~(\ref{k_mode}). The absence of $\dot{\cal E}^{\ell\m}(t)$, however, takes its origin in a much more fundamental property of the fluid dynamics, and it can be justified independently of the mode representation: the fluid dynamics is {\it time-reversal invariant}. This invariance would be violated in the presence of a dissipation mechanism such as viscosity, but throughout this work we assume that no such mechanism is at play. In our new point of view, the time-derivative expansion of Eq.~(\ref{Q_vs_E_time}) is adopted as a fundamental starting point, whose validity is wholly independent of an underlying mode description. And while $k_\ell$ and $\ddot{k}_\ell$ could be calculated on the basis of the mode representation of Eq.~(\ref{k_mode}), we choose, in our mode-less description of dynamical tides, to compute them directly. The methods to achieve this are described in Sec.~\ref{sec:newtonian}. We argue that our mode-less methods offer a practical advantage over the traditional way: To obtain $k_\ell$ and $\ddot{k}_\ell$ we have to integrate a small system of ordinary differential equations, and there is no need to solve an eigenvalue problem for (potentially) a large number of normal modes.

We display our results for polytropic stellar models in Table~\ref{tab:love}.  The static Love numbers of polytropes are of course well known \cite{brooker-olle:55}, and the numbers listed in Table~\ref{tab:love} merely reproduce well-established results (see, for example, Table 2.3 of Poisson and Will \cite{poisson-will:14}). To the best of our knowledge, the dynamic Love numbers were never computed before with the methods proposed here, for any equation of state.

\subsection{Extension of the time-derivative expansion}

But doesn't our mode-less description of dynamical tides, based on the time-derivative expansion of Eq.~(\ref{Q_vs_E_time}), come far short of capturing the approach to resonance that is provided free-of-charge by the mode description? And doesn't this approach to resonance provide the very reason to incorporate dynamical tides in a model of gravitational waves emitted by a neutron-star inspiral? The answer to both questions is of course in the affermative. But what was lost can be recovered. As we shall now argue, there is a way to regain the upper hand by extending the realm of validity of the time-derivative expansion.

Our key observation is that the approximation $\A_{n\ell} = (1-\omega^2/\omega_{n\ell}^2)^{-1} = 1 + \omega^2/\omega_{n\ell} + \cdots$ that is lurking behind Eq.~(\ref{Q_vs_E_time}) can be undone after the fact. Suppose, as is the case in a quasi-circular inspiral, that the tidal moment ${\cal E}^{\ell\m}(t)$ is proportional to $e^{-i\m\Omega t}$, so that it oscillates with a frequency $\m\Omega$, where $\Omega$ is the binary's orbital frequency [Eq.~(\ref{E_circular})]. Then $\ddot{\cal E}^{\ell\m} = -(\m \Omega)^2 {\cal E}^{\ell\m}$ and Eq.~(\ref{Q_vs_E_time}) can be re-expressed as [Eq.~(\ref{Q_vs_E_circ})] 
\begin{equation}
{\cal Q}^{\ell\m}(t) = -\frac{2(\ell-2)!}{(2\ell-1)!!}\, k_\ell \Gamma_\ell^\m\, G^{-1} R^{2\ell+1}\, {\cal E}^{\ell\m}(t)
\label{Q_vs_E_amp}
\end{equation}
with [Eq.~(\ref{Gamma_def})] 
\begin{equation}
\Gamma^\m_{\ell} := 1 + \frac{(\m \Omega)^2}{GM/R^3} \frac{\ddot{k}_\ell}{k_\ell} + \cdots.
\end{equation}
Equation (\ref{Q_vs_E_amp}) with $\Gamma^\m_\ell = 1$ is the usual relationship between mass and tidal multipole moments in the regime of static tides. The additional factor $\Gamma^\m_\ell$ supplies the correction that comes from the dynamical aspects of the tidal interaction, and as it is written here, it is subjected to a low-frequency approximation. To extend its realm of validity we simply re-express $\Gamma^\m_\ell$ as [Eq.~(\ref{Gamma_improved})] 
\begin{equation}
\Gamma^\m_\ell \simeq \biggl[ 1 - \frac{(\m \Omega)^2}{GM/R^3} \frac{\ddot{k}_\ell}{k_\ell} \biggr]^{-1}, 
\label{Gamma_resum} 
\end{equation}
and allow the expression within brackets to become substantially smaller than unity. In this way, the approach to resonance captured by the amplification factor $\A_{n\ell} = (1-\omega^2/\omega_{n\ell}^2)^{-1}$ is successfully recreated in a mode-less description of dynamical tides.

When we insert Eq.~(\ref{Gamma_resum}) within Eq.~(\ref{Q_vs_E_amp}) and take a Fourier transform, we obtain Eq.~(\ref{Q_vs_E_freq}) with a frequency-domain Love number given by [Eqs.~(\ref{omega_star}) and (\ref{k_resum2})]
\begin{equation}
\tilde{k}_\ell(\omega) = \frac{\tilde{k}_\ell(0)}{1 - \omega^2/\omega^2_{*\ell}}, \qquad 
\omega^2_{*\ell} := \frac{GM}{R^3} \frac{k_\ell}{\ddot{k}_\ell},
\label{k_onemode} 
\end{equation}
where $\omega_{*\ell}$ is an effective frequency defined in terms of the static and dynamic Love numbers. We notice that Eq.~(\ref{k_onemode}) is formally identical to Eq.~(\ref{k_mode}) when the sum over modes is truncated to a single term. Furthermore, in Sec.~\ref{subsec:fmode} --- refer to Table~\ref{tab:ratio} --- we show that $\omega_{*\ell}$ is numerically very close to the frequency $\omega_{0\ell}$ of the star's {\it fundamental mode} (or $f$-mode) of vibration, labelled by $n=0$. Our conclusion from these observations is that the extension of the time-derivative expansion formulated in Eq.~(\ref{Gamma_resum}) should provide a description of dynamical tides that compares very well in accuracy with a mode-sum representation truncated to the $f$-mode. Because the $f$-mode comes with mode functions with the least number of radial nodes, it is expected to provide by far the largest contribution to $\tilde{k}_\ell(\omega)$, and therefore an excellent approximation to it. Our mode-less description of dynamical tides will do just as well.\footnote{The quality of the approximations can be assessed by plotting the various versions of $\tilde{k}_\ell(\omega)$ on the same graph. We carried out this exercise for a $n=1$ polytrope, and plotted $\tilde{k}_\ell(\omega)$ for (i) the approximation of Eq.~(\ref{k_onemode}), (ii) the $f$-mode truncation of Eq.~(\ref{k_mode}), and (iii) an exact representation equivalent to a sum over all modes. In a range of frequencies extending from $\omega = 0$ to almost the $f$-mode frequency, over which $\tilde{k}_\ell(\omega)$ increases by more than an order of magnitude, the plots look completely identical. The one-mode approximation and $f$-mode truncation are therefore extremely accurate. We chose not to include these plots here, because there is literally nothing to see.} 

Our model of dynamical tides, based on Eq.~(\ref{k_onemode}), does not capture all aspects of the dynamical regime; it shares the same limitations as the $f$-mode truncation of the spectral representation. For example, these models leave out the resonant excitation of $g$-modes, which come with frequencies that are lower than the $f$-mode frequency, and which can indeed become resonant during an inspiral. (Our simplistic modeling of stellar structure, based on a polytropic equation of state, exclude the presence of $g$-modes. But they are present in complete models of neutron stars.) In this case, however, the limitation is not severe, as the dynamical impact of resonant $g$-modes was shown \cite{lai:94, reisenegger-goldreich:94} to be negligible in binary inspirals; the modes are resonant, but they come with exceedingly small overlap integrals. A more serious limitation presents itelf in the case of resonant {\it inertial modes}, which manifest themselves in inspirals implicating rotating stars \cite{flanagan-racine:07, poisson:20a}. In such cases a number of dynamically significant resonances can occur, and these are not captured by the model of Eq.~(\ref{k_onemode}), nor by an $f$-mode truncation of the spectral representation. A complete description of dynamical tides that includes the effects of significant resonances would require additional ingredients. Our model and the $f$-mode truncation are not so ambitious; they are meant to capture the dynamical effects associated with the approach to a resonance that is never actually reached.

In the purely Newtonian context considered thus far, the mode-less approach to dynamical tides does not come with a decisive advantage over the $f$-mode truncation, although we did argue that it produces a saving in computational tasks --- there is no need to solve an eigenvalue problem to find the modes. The advantage, however, comes loud and clear in the context of general relativity, in view of the obstacles reviewed in Sec.~\ref{subsec:obstacles}. We turn to this next.  

\subsection{Mode-less description of dynamical tides: General relativity}

The time-derivative expansion of Eq.~(\ref{Q_vs_E_time}) applies also in general relativity. The equation, however, comes with a subtle interpretation that was fully articulated in Ref.~\cite{poisson:21a}. We go briefly over these points of interpretation here, and refer the reader to the earlier work for a more complete discussion.

The most delicate issue in establishing the validity of Eq.~(\ref{Q_vs_E_time}) in general relativity is to provide a proper definition for the mass multipole moments of an individual body in a dynamical spacetime that may contain any number of compact bodies. The solution proposed in Ref.~\cite{poisson:21a} takes as input a situation in which the {\it mutual gravity} between the bodies is sufficiently weak to be adequately described by a post-Newtonian expansion of the metric, while the {\it individual gravity} of each body can be arbitrarily strong. When viewed from close by, in a spacetime described in full general relativity, each body is revealed as an extended object deformed by tidal forces. But when it is viewed from far away, from the vantage point of the post-Newtonian spacetime, the body appears as a skeletonized object with a specific multipole structure, moving on some world line. A matching of these two descriptions of the same spacetime reconciles the points of view and permits a determination of the tidal moments ${\cal E}^{\ell\m}(t)$ and mass multipole moments ${\cal Q}^{\ell\m}(t)$. To leading (Newtonian) order in the post-Newtonian expansion of the mutual gravity, we obtain Eq.~(\ref{Q_vs_E_time}), with the important proviso that $k_\ell$ and $\ddot{k}_\ell$ must now be computed in full general relativity. It is in this sense that Eq.~(\ref{Q_vs_E_time}) applies in general relativity. Corrections of the first post-Newtonian order were also calculated in Ref.~\cite{poisson:21a}, and corrections of higher order can be added as they become available. 

\begin{figure}
\includegraphics[width=0.49\linewidth]{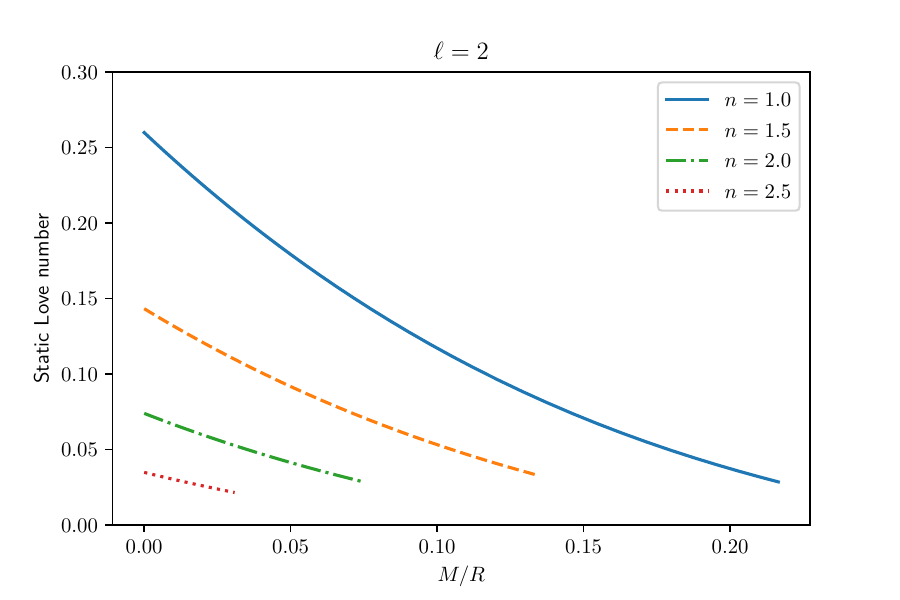}
\includegraphics[width=0.49\linewidth]{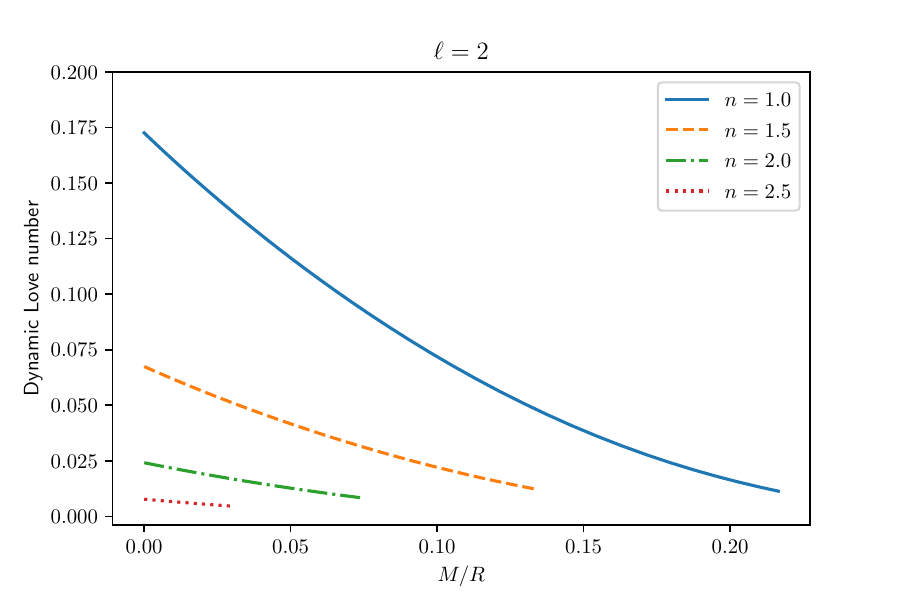}
\caption{Static and dynamic Love numbers for $\ell =2$, computed for relativistic stellar models with an equation of state $p = K \rho^{1+1/n}$, where $n = \{1.0, 1.5, 2.0, 2.5\}$. Left panel: static Love number $k_\ell$. Right panel: dynamic Love number $\ddot{k}_\ell$. Each Love number is plotted as a function of the stellar compactness $M/R$. The curve ends at the configuration of maximum mass, beyond which the equilibrium sequence is dynamically unstable.}  
\label{fig:loveL2} 
\end{figure} 

A Newtonian-order approximation to a fully relativistic result may seem somewhat crude, but it is important to recall that in the context of a binary dynamics implicating compact objects, the Newtonian-order Eq.~(\ref{Q_vs_E_time}) actually gives rise to a tidal interaction at the fifth post-Newtonian order; corrections of the first post-Newtonian order translate to an interaction at the sixth post-Newtonian order. The low-order results of Ref.~\cite{poisson:21a} therefore do a good job of describing the tidal dynamics at high post-Newtonian orders; alternative treatments can be found in Refs.~\cite{vines-flanagan:13, henry-faye-blanchet:20a}. To see where the boost in post-Newtonian orders is coming from, consider the leading, quadrupole term in the tidal interaction. The tidal moment scales as ${\cal E} \sim G M'/d^3$, in which $M'$ is the companion's mass and $d$ is the orbital separation, and Eq.~(\ref{Q_vs_E_time}) produces ${\cal Q} \sim M' R^5/d^3$. The tidal contribution to the gravitational force is then $F_{\rm tide} \sim G M' {\cal Q}/d^4 \sim G M^{\prime 2} R^5/d^7$, and when this is divided by the point-mass contribution $G M M'/d^2$, we obtain the ratio $(M'/M) (R/d)^5$. For a compact object with a radius $R \sim GM/c^2$, the ratio becomes $(M'/M) (GM/c^2 d)^5$, and the tidal force takes the form of a correction of the fifth post-Newtonian order. 

We therefore have that Eq.~(\ref{Q_vs_E_time}) provides a meaningful description of dynamical tides in general relativity, provided that $k_\ell$ and $\ddot{k}_\ell$ are computed in the relativistic setting. We perform such a calculation in Sec.~\ref{sec:GR}. We display a sample of our results in Fig.~\ref{fig:loveL2}, in which $k_\ell$ and $\ddot{k}_\ell$ are plotted as functions of stellar compactness $M/R$ for polytropic stellar models. We see that for $M/R \to 0$, the relativistic numbers agree with the Newtonian values listed in Table~\ref{tab:love}. We see also that the Love numbers decrease with increasing $M/R$, reaching a minimum when the equilibrium sequence comes to an end at the configuration of maximum mass. 

The bold step of going from Eq.~(\ref{Q_vs_E_time}) to Eq.~(\ref{Q_vs_E_amp}), with an amplification factor expressed as in Eq.~(\ref{Gamma_resum}), can also be taken in general relativity. This gives us a most promising starting point for a description of dynamical tides in relativistic mechanics. We hope to have conveyed a sense that this description comes with a clear and sound foundation, and that it will produce a practical and reliable model for neutron-star inspirals. With this our narrative ends, and the remaining portions of the paper contain all the technical details.  

\subsection{Organization of the paper} 

We begin our technical developments in Sec.~\ref{sec:newtonian}, where the static and dynamic Love numbers are defined and computed in Newtonian fluid mechanics and gravitation. Our fundamental starting point is the description of dynamical tides provided by Eq.~(\ref{Q_vs_E_time}), which relates the mass multipole moments of a tidally deformed body to the tidal multipole moments and their time derivatives. The perturbation equations that describe the fluid deformation are derived systematically through second order in the time-derivative expansion, and the system of ordinary differential equations is integrated for polytropic stellar models based on the equation of state $p = K \rho^{1+1/n}$.

In Sec.~\ref{sec:fmode} we return to the mode picture, and provide $\tilde{k}_\ell(\omega)$ --- and therefore $k_\ell$ and $\ddot{k}_\ell$ --- with a spectral representation in terms of the star's normal modes of vibration; this produces Eq.~(\ref{k_mode}), above. We describe its truncation to an $f$-mode approximation, and introduce a low-frequency approximation that reproduces Eq.~(\ref{Q_vs_E_time}). We then construct the extension of Eq.~(\ref{Gamma_resum}) and compare the resulting description of dynamical tides with the $f$-mode approximation. 

In Sec.~\ref{sec:GR} we turn to a computation of $k_\ell$ and $\ddot{k}_\ell$ in full general relativity. The calculation is based on a careful matching of the exterior and interior metrics of a tidally deformed body at the surface. The exterior metric is imported from Ref.~\cite{poisson:21a}, and the interior metric is obtained by integrating the equations that govern the tidal deformation of a relativistic perfect fluid. We again select the polytropic equation of state $p = K \rho^{1+1/n}$, and our results are summarized in Fig.~\ref{fig:loveL2} above, Figs.~\ref{fig:loveL3}, \ref{fig:loveL4}, and \ref{fig:loveL5} below, as well as Tables~\ref{tab:k} and \ref{tab:kddot} below.  

A discussion of our numerical methods is relegated to Appendix~\ref{sec:numerical}. 

\section{Static and dynamic Love numbers in Newtonian gravity} 
\label{sec:newtonian}

Our task in this section is to define and compute, within the framework of Newtonian fluid mechanics and gravitation, the static and dynamic Love numbers of a spherical star deformed under the action of a time-dependent tidal field. We begin in Sec.~\ref{subsec:potential} with a description of a time-changing tidal field and the associated body's response as measured in the exterior gravitational field. We introduce the fluid equations in Sec.~\ref{subsec:governing}, and derive from them a system of perturbation equations in Sec.~\ref{subsec:perturbation}. In Sec.~\ref{subsec:polytrope} we specialize these equations to the specific case of polytropic stellar models, and in Sec.~\ref{subsec:love} we obtain the Love numbers listed in Table~\ref{tab:love}.
An introduction to the tidal deformation of fluid bodies can be found in Sec.~2.4 of Poisson and Will \cite{poisson-will:14}.  

\subsection{Tidal potential and body's response}
\label{subsec:potential} 

We consider a (nonrotating) body of mass $M$ and radius $R$, and imagine that it is made up of a perfect fluid. The body is spherical when isolated, but a deformation is created when remote objects exert tidal forces. We wish to characterize this deformation in terms of Love numbers. We imagine that the forces are sufficiently small that the deformation can be adequately described within the framework of first-order perturbation theory. We further imagine that the tidal field varies slowly with time (with an external, orbital time scale much longer than the internal, hydrodynamical time scale), so that the tidal response can be expressed as a time-derivative expansion. 

Assuming that each remote object is at a large distance from the reference body, we expand the external gravitational potential $U^{\rm ext}$ --- the potential created by the objects --- in powers of $r/d \ll 1$, in which $r := |\bm{x}|$ is the distance from the body's center-of-mass, and $d$ is the typical distance to an external object. The $\ell$-th order term in the Taylor expansion of the external potential is
\begin{equation}
U_\elltide(t,\bm{x}) = -\frac{1}{(\ell-1)\ell}\, r^\ell\, {\cal E}_L \Omega^L,
\label{tidal_potential1} 
\end{equation} 
where
\begin{equation}
{\cal E}_L(t) := -\frac{1}{(\ell-2)!} \partial_L U^{\rm ext} \biggr|_{\bm{x}=\bm{0}}
\label{EL_def} 
\end{equation}
is the tidal multipole moment and $\bm{\Omega} := \bm{x}/r = [\sin\theta\cos\phi, \sin\theta\sin\phi, \cos\theta]$ is the radial unit vector. The multi-index $L$ contains a number $\ell$ of individual indices, so that $\partial_L U^{\rm ext} := \partial_{a_1} \partial_{a_2} \cdots \partial_{a_\ell} U^{\rm ext}$, and we use the notation $\Omega^L := \Omega^{a_1} \Omega^{a_2} \cdots \Omega^{a_\ell}$; summation over repeated indices is implied. The moment ${\cal E}_L(t)$ is symmetric and tracefree in all its indices. It should be noted that the normalization of the tidal moments in Eq.~(\ref{EL_def}) follows the conventions of Binnington and Poisson \cite{binnington-poisson:09} (which came from Zhang \cite{zhang:86}); it differs from the normalization adopted in Poisson and Will \cite{poisson-will:14}. 

We write 
\begin{equation}
{\cal E}_L \Omega^L = \sum_{\m=-\ell}^\ell {\cal E}^{\ell\m}\, Y^{\ell\m}(\theta,\phi),
\label{EL_vs_Eell} 
\end{equation}
where $Y^{\ell\m}(\theta,\phi)$ are the standard spherical harmonics. By virtue of being symmetric and tracefree, the tensor ${\cal E}_L$ possesses a number $2\ell+1$ of independent components, and Eq.~(\ref{EL_vs_Eell}) puts them in a one-to-one correspondence with the $2\ell+1$ coefficients ${\cal E}^{\ell\m}$. Equation (\ref{tidal_potential1}) can be rewritten as
\begin{equation}
U_\elltide(t,\bm{x}) = -\frac{1}{(\ell-1)\ell}\, {\cal E}^{\ell\m}(t)\, r^\ell Y^{\ell\m}(\theta,\phi),   
\label{tidal_potential2} 
\end{equation} 
where we leave the summation over $\m$ implicit. 

The tidal forces exerted by the remote objects deform the body from its original spherical state. The deformation is measured by the mass multipole moment
\begin{equation}
{\cal Q}^L(t) := \int \rho\, x^\stf{L}\, dV,
\label{QL_def}
\end{equation}
where $\rho$ is the body's mass density and $x^\stf{L}$ is the tracefree projection of $x^L := x^{a_1} x^{a_2} \cdots x^{a_\ell}$. The corresponding term in the body's potential is [Eqs.~(1.149), (1.156), and (1.157) of Poisson and Will] 
\begin{equation}
U_\ellmass(t,\bm{x}) = \frac{(-1)^\ell}{\ell!}\, {\cal Q}^L \partial_L \frac{1}{r}
= \frac{(2\ell-1)!!}{\ell!}\, r^{-(\ell+1)} {\cal Q}_L \Omega^L.
\label{resp_potential1}
\end{equation}
In analogy with Eq.~(\ref{EL_vs_Eell}) we write 
\begin{equation}
{\cal Q}_L \Omega^L = \sum_{\m=-\ell}^\ell {\cal Q}^{\ell\m} Y^{\ell\m}(\theta,\phi),   
\label{QL_vs_Qell}
\end{equation}
and re-express Eq.~(\ref{resp_potential1}) as 
\begin{equation}
U_\ellmass(t,\bm{x}) = \frac{(2\ell-1)!!}{\ell!}\, {\cal Q}^{\ell\m}(t)\, r^{-(\ell+1)} Y^{\ell\m}(\theta,\phi), 
\label{resp_potential2}
\end{equation}
where we again omit the summation over $\m$. 

The body's deformation, measured by the mass multipole moments ${\cal Q}^{\ell\m}$, is determined by the tidal field, which is characterized by the tidal multipole moments ${\cal E}^{\ell\m} $. In a context in which the forces and deformation are small, the relationship between the mass and tidal moments shall be linear, and we make this approximation throughout this work --- nonlinearities were considered in Ref.~\cite{poisson:21a}. In a context in which the tidal forces vary slowly with time, we can further express the relationship as a time-derivative expansion. We thus write [Eq.~(\ref{Q_vs_E_time})]
\begin{equation}
G{\cal Q}^{\ell\m} (t) = -\frac{2(\ell-2)!}{(2\ell-1)!!} R^{2\ell+1} 
\biggl[ k_\ell\, {\cal E}^{\ell\m} (t) - \ddot{k}_\ell \frac{R^3}{GM}\, \ddot{\cal E}^{\ell\m} (t)
+ \cdots \biggr], 
\label{Q_vs_E_dyn} 
\end{equation}
with overdots on ${\cal E}^{\ell\m}$ indicating differentiation with respect to time. The overall numerical factor on the right-hand side is conventional, and the factor of $R^{2\ell+1}$ ensures that the static Love number $k_\ell$ is dimensionless. The additional factor of $R^3/(GM)$ in front of the second-derivative term has the dimension of a time squared and therefore compensates dimensionally for the time derivatives of the tidal moment; its presence ensures that the dynamic Love number $\ddot{k}_\ell$ is also dimensionless. The definition of $\ddot{k}_\ell$ adopted in Eq.~(\ref{Q_vs_E_dyn}) differs by a minus sign from the one introduced in Ref.~\cite{poisson:21a}. In the new convention the dynamic Love numbers will turn out to be positive. 

As we explained in Sec.~\ref{sec:intro}, Eq.~(\ref{Q_vs_E_dyn}) should be viewed as an expansion of the mass moment in powers of the ratio
\begin{equation} 
\epsilon := \frac{\mbox{internal, hydrodynamical time scale}}{\mbox{external, orbital time scale}} \ll 1.
\label{epsilon_def} 
\end{equation} 
In principle the time-derivative expansion could be extended to higher powers of $\epsilon$, but the dynamical correction to the mass moment will be dominated by the first nonvanishing power, $\epsilon^2$. A notable fact is the absence of a correction of order $\epsilon$. The absence of odd terms in the expansion is dictated by the time-reversal invariance of the fluid equations, which will be introduced below. This invariance could be broken by the introduction of dissipation within the fluid, for example in the form of viscous heating. In the context of a perfect fluid, however, no dissipation takes place, and the physics is time-reversal invariant.

Combining Eqs.~(\ref{tidal_potential2}), (\ref{resp_potential2}), and Eq.~(\ref{Q_vs_E_dyn}), we find that the $\ell$-pole contribution to the gravitational potential is given by 
\begin{equation}
\delta U = -\frac{1}{(\ell-1)\ell} \Biggl[
\biggl( r^\ell + 2k_\ell \frac{R^{2\ell+1}}{r^{\ell+1}} \biggr) {\cal E}^{\ell\m}
- 2 \ddot{k}_\ell \frac{R^3}{GM} \frac{R^{2\ell+1}}{r^{\ell+1}} \ddot{\cal E}^{\ell\m} \Biggr]
Y^{\ell\m}(\theta,\phi). 
\label{Uext_dyn} 
\end{equation}
Our goal in the remainder of this section is to elaborate a method to compute $k_\ell$ and $\ddot{k}_\ell$. To achieve this we must now turn to the body's interior. 

\subsection{Governing equations} 
\label{subsec:governing} 

We take the body to consist of a perfect fluid. The equations that govern the body's structure are (i) Euler's equation
\begin{equation}
\partial_t v_a + v^b \nabla_b v_a = \nabla_a U - \rho^{-1} \nabla_a p,
\label{euler}
\end{equation} 
where $v^a$ is the velocity field, $\rho$ the mass density, $p$ the pressure, and $U$ the gravitational potential, (ii) the continuity equation
\begin{equation}
\partial_t \rho + \nabla_a (\rho v^a) = 0,
\label{continuity}
\end{equation}
and (iii) Poisson's equation
\begin{equation}
\nabla^2 U = -4\pi G \rho.
\label{poisson}
\end{equation}
We work in arbitrary coordinates $x^a$, with a metric $g_{ab}$ and a compatible covariant derivative $\nabla_a$. The equations are supplemented with an equation of state, which we take to be of a simple barotropic form, $p = p(\rho)$. This allows us to define a specific enthalpy $h(\rho)$ via $dh = \rho^{-1}\, dp$. With this variable the equation of state can be written in the parametric form $\rho = \rho(h)$, $p = p(h)$.

When the fluid is in a state of static equilibrium we have that $v^a = 0$, and Euler's equation reduces to
\begin{equation}
\rho^{-1} \nabla_a p = \nabla_a h = \nabla_a U.
\label{euler_eq}
\end{equation}
The continuity equation is then trivially satisfied, and Poisson's equation stays unchanged.

When the equilibrium is spherically symmetric, the equations reduce to
\begin{equation}
m' = 4\pi r^2 \rho, \qquad
U' = -Gm/r^2, \qquad
h' = -G m/r^2,
\label{sph_eq}
\end{equation}
where $m(r)$ is the mass inside a sphere of radius $r$; a prime indicates differentiation with respect to $r$. The potential and specific enthalpy are related by $h(r) = U(r) - GM/R$, where $M := m(r=R)$ is the body's total mass and $r = R$ marks the stellar surface; this ensures that $h$ properly vanishes at the surface.  

\subsection{Perturbation equations}
\label{subsec:perturbation}

We now perturb the spherical equilibrium by introducing a time-dependent tidal field produced by remote objects. We denote by $\delta \rho$, $\delta p$, $\delta h$, and $\delta U$ the Eulerian perturbations in density, pressure, specific enthalpy, and gravitational potential, respectively. The fluid's velocity field no longer vanishes, and we write it as $\delta v^a$. 

Linearization of the governing equations produces
\begin{equation}
\partial_t \delta v_a = \nabla_a(\delta U - \delta h), \qquad
\partial_t \delta \rho + \nabla_a (\rho\, \delta v^a) = 0, \qquad
\nabla^2 \delta U = -4\pi G \delta\rho.
\label{pert_dyn1}
\end{equation}
The equation of state implies
\begin{equation}
\delta \rho = \frac{d\rho}{dh}\, \delta h = \frac{\rho'}{h'}\, \delta h
= -\frac{r^2 \rho'}{Gm}\, \delta h
\end{equation}
and $\delta p = \rho\, \delta h$. 

The perturbed Euler equation reveals that $\delta v_a$ is a gradient field,
\begin{equation}
\delta v_a = \nabla_a \delta \psi,
\end{equation}
where $\delta \psi$ is some potential. The equation becomes
\begin{equation}
\partial_t \delta \psi =  \delta U - \delta h, 
\label{pert_dyn2}
\end{equation}
and
\begin{equation}
\partial_t \delta \rho + \nabla_a \rho\, \nabla^a \delta \psi + \rho \nabla^2 \delta \psi = 0
\label{cont_dyn}
\end{equation} 
is the new statement of mass conservation. 

Taking our cue from Eq.~(\ref{Uext_dyn}), we expand the perturbation variables as
\begin{subequations}
\begin{align}
\delta \rho &= \biggl( \rho_\ell^0\, {\cal E}^{\ell\m}
+ \rho_\ell^2 \frac{R^3}{GM} \ddot{\cal E}^{\ell\m} \biggr) Y^{\ell\m}, \\
\delta h &= \biggl( h_\ell^0\, {\cal E}^{\ell\m}
+ h_\ell^2 \frac{R^3}{GM} \ddot{\cal E}^{\ell\m} \biggr) Y^{\ell\m}, \\
\delta U &= \biggl( U_\ell^0\, {\cal E}^{\ell\m}
+ U_\ell^2 \frac{R^3}{GM} \ddot{\cal E}^{\ell\m} \biggr) Y^{\ell\m},\\
\delta \psi &= \psi_\ell^1 \frac{R^3}{GM} \dot{\cal E}^{\ell\m}\, Y^{\ell\m}, 
\end{align}
\end{subequations}
where the coefficients $\rho_\ell^0$, $\rho_\ell^2$, and so on, are functions of $r$. The scaling of $\delta \psi$ with $\dot{\cal E}^{\ell\m}$ is justified by the observations that (i) the velocity perturbation $\delta v_a = \nabla_a \delta \psi$ must vanish in the static limit, (ii) time-reversal invariance dictates the absence of even time derivatives, and (iii) the first derivative provides the leading contribution.

Making the substitutions within the governing equations, we find that the zeroth-order radial functions satisfy
\begin{equation}
h_\ell^0 = U_\ell^0,
\label{h0_vs_U0}
\end{equation}
and
\begin{equation}
r^2 \frac{d^2 U^0_\ell}{dr^2} + 2r \frac{dU^0_\ell}{dr} - \ell(\ell+1) U^0_\ell + 4\pi G r^2 \rho^0_\ell = 0.
\label{U0_eq}
\end{equation}
The equation of state implies that 
\begin{equation}
\rho_\ell^0 = -\frac{r^2 \rho'}{Gm} h_\ell^0, 
\label{rho0_vs_h0} 
\end{equation}
and we arrive at a self-contained set of equations for $h^0_\ell$ and $U^0_\ell$. 

At the next order we have that the perturbed Euler equation (\ref{pert_dyn2}) produces
\begin{equation}
h_\ell^2 = U_\ell^2 - \psi_\ell^1,
\label{h2_vs_U2}
\end{equation}
and the perturbed continuity equation (\ref{cont_dyn}) gives rise to
\begin{equation}
r^2 \frac{d^2 \psi_\ell^1}{dr^2} + \biggl( 2 + \frac{r\rho'}{\rho} \biggr) r \frac{d \psi_\ell^1}{dr}
- \ell(\ell+1) \psi_\ell^1 - \frac{M}{R^3} \frac{r^4 \rho'}{m\rho}\, U_\ell^0 = 0;
\label{psi2_eq} 
\end{equation}
to arrive at this result we made use of Eqs.~(\ref{h0_vs_U0}) and (\ref{rho0_vs_h0}). The perturbed Poisson equation yields
\begin{equation}
r^2 \frac{d^2 U_\ell^2}{dr^2} + 2r \frac{d U_\ell^2}{dr}
- \biggl[ \ell(\ell+1) + \frac{4\pi r^4 \rho'}{m} \biggr] U_\ell^2
+ \frac{4\pi r^4 \rho'}{m}\, \psi_\ell^1 =0, 
\label{U2_eq} 
\end{equation}
where we made use of the equation of state to express $\rho_\ell^2$ in terms of $h_\ell^2$, and then Eq.~(\ref{h2_vs_U2}) to write this in terms of $U^2_\ell$ and $\psi^1_\ell$. We again have a self-contained set of equations. 

The occurrence of $\rho'$ in the perturbation equations makes them awkward to integrate when $\rho' \to -\infty$ at the body's surface, which happens for some equations of state. An example is the polytropic form adopted below, $p = K \rho^{1 + 1/n}$, when $n < 1$. It is known that Eq.~(\ref{U0_eq}) can be tamed by implementing the transformation $U^0_\ell = (Gm/r) f^0_\ell$, where $f^0_\ell$ is a new dependent variable. The equation then becomes Clairaut's equation, displayed as Eq.~(2.230) in Poisson and Will \cite{poisson-will:14}. It should be possible to tame the remaining equations by adopting new variables in lieu of $\psi^1_\ell$ and $U^2_\ell$, but we have not pursued this option. In our computations below we shall simply avoid cases where $\rho'$ is unbounded at the body's surface. 

It can be shown that regular solutions to the perturbation equations behave as $r^\ell$ when $r \to 0$. To account for this behavior it is helpful to set 
\begin{equation}
U_\ell^0 = r^\ell\, e_\ell^0, \qquad
\psi_\ell^1 = r^\ell\, e_\ell^1, \qquad
U_\ell^2 = r^\ell\, e_\ell^2, 
\end{equation}
with $e_\ell^j$ denoting the new set of perturbation variables. The perturbation equations become
\begin{subequations}
\label{e_eqns} 
\begin{align}
0 &= r^2 \frac{d^2 e_\ell^0}{dr^2} + 2(\ell+1) r \frac{d e_\ell^0}{dr}
- \frac{4\pi r^4\rho'}{m}\, e_\ell^0, \\
0 &= r^2 \frac{d^2 e_\ell^1}{dr^2} + \biggl[ 2(\ell+1) + \frac{r\rho'}{\rho} \biggr] \frac{d e_\ell^1}{dr} 
+ \ell \frac{r\rho'}{\rho}\, e_\ell^1 - \frac{M}{R^3} \frac{r^4 \rho'}{m\rho}\, e_\ell^0, \\
0 &= r^2 \frac{d^2 e_\ell^2}{dr^2} + 2(\ell+1) r \frac{d e_\ell^2}{dr}
- \frac{4\pi r^4\rho'}{m}\, \bigl( e_\ell^2 - e_\ell^1 \bigr).
\end{align}
\end{subequations}

The interior solutions for $e_\ell^0$ and $e_\ell^2$ are to be matched at $r=R$ to the exterior solutions provided by Eq.~(\ref{Uext_dyn}). We have that 
\begin{subequations}
\label{e_ext}
\begin{align} 
e^0_\ell \bigr|_{\rm ext} &= -\frac{1}{(\ell-1)\ell} \Bigl[ 1 + 2 k_\ell (R/r)^{2\ell+1} \Bigr],  \\
e^2_\ell \bigr|_{\rm ext} &= \frac{1}{(\ell-1)\ell} (2 \ddot{k}_\ell) (R/r)^{2\ell+1}.   
\end{align} 
\end{subequations}
The matching determines the Love numbers $k_\ell$ and $\ddot{k}_\ell$.

\subsection{Implementation for a polytrope} 
\label{subsec:polytrope} 

The integration of Eq.~(\ref{e_eqns}) requires prior knowledge of the functions $\rho(r)$ and $m(r)$, and this is obtained by integrating Eqs.~(\ref{sph_eq}) for the stellar structure. When the fluid is barotropic it is convenient to reformulate the structure equations by letting $h$ (or its substitute $\vartheta$, as we shall see presently) become the independent variable \cite{lindblom:92}. We describe such a formulation here.

We rescale the variables according to
\begin{equation}
h = h_c\, \vartheta, \qquad
m = \frac{4\pi}{3} \rho_c r^3\, \chi, \qquad
r^2 = r_0^2\, \zeta,
\label{structure_variables} 
\end{equation}
where $\vartheta$, $\chi$, and $\zeta$ are new dimensionless variables; $h_c := h(r=0)$ is the central value of the specific enthalpy, $\rho_c := \rho(r=0)$ is the central value of the density, and $r_0^2 := 3 h_c/(2\pi G \rho_c)$. The structure equations become
\begin{equation}
\frac{d\vartheta}{d\zeta} = -\chi, \qquad
\frac{d\chi}{d\zeta} = \frac{3}{2\zeta} (\rho/\rho_c - \chi), 
\end{equation}
and they are rewritten in the form
\begin{equation}
\frac{d\zeta}{d\vartheta} = -\frac{1}{\chi}, \qquad
\frac{d\chi}{d\vartheta} = -\frac{3}{2\zeta \chi} (\rho/\rho_c - \chi),
\label{structure_theta}
\end{equation}
with $\vartheta$ playing the role of independent variable. The equations are integrated from $\vartheta = 1$ with boundary conditions $\zeta(\vartheta=1) = 0$, $\chi(\vartheta=1) = 1$, up to $\vartheta = 0$, which marks the body's surface. The surface values $\zeta_s := \zeta(\vartheta = 0)$ and $\chi_s := \chi(\vartheta=0)$ are obtained from the integration, and from them we get
\begin{equation}
M = \frac{4\pi}{3} \rho_c R^3\, \chi_s, \qquad
R = r_0\, \zeta_s^{1/2},
\end{equation}
respectively the body's total mass and radius.

The integration of the structure equations require the specification of an equation of state. We make the simple choice of a polytropic form, 
\begin{equation}
p = K \rho^{1+1/n},
\end{equation}
where $K$ and $n$ are constants. It is easy to show that in this case, $h = (n+1) K \rho^{1/n}$, so that $\rho = \rho_c \vartheta^n$.  

It is helpful to reformulate the perturbation equations as a first-order system, by introducing the auxiliary variables
\begin{equation}
v_\ell^0 := r \frac{d e_\ell^0}{dr}, \qquad
v_\ell^1 := r \frac{d e_\ell^1}{dr}, \qquad
v_\ell^2 := r \frac{d e_\ell^2}{dr}.
\label{e_eqns_dyn}
\end{equation}
Making the substitutions in Eq.~(\ref{e_eqns}), and rewriting in terms of the dimensionless variables introduced previously, we find that the perturbation equations become  
\begin{subequations}
\label{v_eqns_dyn} 
\begin{align}
0 &= r \frac{d v_\ell^0}{dr} + (2\ell+1) v_\ell^0 + 6 n \zeta \vartheta^{n-1}\, e_\ell^0,
\label{v0_eq} \\
0 &= r \frac{d v_\ell^1}{dr} + \bigg( 2\ell+1 - 2n \frac{\zeta \chi}{\vartheta} \biggr) v_\ell^1
- 2 n \ell \frac{\zeta \chi}{\vartheta}\, e_\ell^1 + 2 n \chi_s \frac{\zeta}{\vartheta}\, e_\ell^0,
\label{v1_eqn} \\
0 &= r \frac{d v_\ell^2}{dr} + (2\ell+1) v_\ell^2 + 6 n \zeta \vartheta^{n-1} (e_\ell^2 - e_\ell^1),
\label{v2_eqn} 
\end{align}
\end{subequations}
where $\chi_s := \chi(\vartheta=0)$ is the surface value of the structure function. The differential operator is now interpreted as 
\begin{equation}
r \frac{d}{dr} = -2\zeta \chi \frac{d}{d\vartheta}.
\end{equation}

The dynamical system comes with two sets of boundary conditions. At the center ($\vartheta = 1$) we impose
\begin{equation}
v_\ell^0 = 0, \qquad v_\ell^1 = 0, \qquad v_\ell^2 = 0;
\end{equation}
these equations follow from the definition of the auxiliary variables. At the surface ($\vartheta = 0$) we impose
\begin{equation}
e_\ell^0 = 1, \qquad
v_\ell^1 + \ell e_\ell^1 - e_\ell^0 = 0, \qquad
v_\ell^2 + (2\ell+1) e_\ell^2 = 0.
\end{equation}
The first equation, $e_\ell^0(\vartheta=0)=1$, reflects a choice of normalization for the solution; we shall call a solution that satisfies this condition a {\it normalized solution}. The second equation ensures that the terms proportional to $\vartheta^{-1}$ in Eq.~(\ref{v1_eqn}) do not cause the solution to blow up on the surface. The third equation follows from Eq.~(\ref{e_ext}).

\subsection{Love numbers}
\label{subsec:love} 

We first describe a method to compute the static Love number $k_\ell$ from the normalized solution to Eq.~(\ref{v0_eq}), which we denote by $\hat{e}_\ell^0$ and $\hat{v}_\ell^0$. Because the equations for $e_\ell^0$ and $v_\ell^0$ are linear, the true solution --- the one matching the exterior solution of Eq.~(\ref{e_ext}) --- will be $e^0_\ell = N_\ell\, \hat{e}^0_\ell$ and $v^0_\ell = N_\ell\, \hat{v}^0_\ell$, for some number $N_\ell$. Let
\begin{equation}
\eta_\ell := \frac{v^0_\ell}{e^0_\ell} \biggr|_{r=R}
= \frac{\hat{v}^0_\ell}{\hat{e}^0_\ell} \biggr|_{r=R} = \hat{v}^0_\ell(r=R).
\end{equation}
According to Eq.~(\ref{e_ext}), this quantity must be given by $\eta_\ell = -(2\ell+1) (2k_\ell)/(1 + 2k_\ell)$. Solving for $k_\ell$, we have that
\begin{equation}
2 k_\ell = -\frac{\eta_\ell}{2\ell+1+\eta_\ell}.
\label{kell_vs_etaell}
\end{equation}
We also find that 
\begin{equation}
N_\ell = -\frac{2\ell+1}{(\ell-1)\ell} \frac{1}{2\ell+1 + \eta_\ell}
\end{equation}
is the correct value for the normalization constant. 

The method is easily extended to produce a value for the dynamic Love number $\ddot{k}_\ell$. Because the entire system of perturbation equations is linear, the same numerical factor $N_\ell$ converts $\hat{e}^2_\ell$, the normalized solution to the system of equations, to the true solution $e_\ell^2$ that matches the external form provided by Eq.~(\ref{e_ext}). This implies that
\begin{equation}
2 \ddot{k}_\ell = -\frac{2\ell+1}{2\ell+1+\eta_\ell}\, \hat{e}_\ell^2(\vartheta=0). 
\end{equation}

A numerical integration of the perturbation equations returns the static and dynamic Love numbers of Table~\ref{tab:love}. We carried out the integrations with a collocation method based on an expansion of all variables in Chebyshev polynomials; the method is detailed in Appendix \ref{sec:numerical}. 

\section{Love numbers in terms of normal modes}
\label{sec:fmode}

In this section we stay within the framework of Newtonian fluid mechanics and gravitation, and construct an alternate description of the tidal deformation of a (nonrotating) body of mass $M$ and radius $R$, in terms of the body's normal modes of vibration. This description allows us to define the frequency-domain Love number $\tilde{k}_\ell(\omega)$ that was first encountered in Eq.~(\ref{Q_vs_E_freq}), and which was expressed as a sum over modes in Eq.~(\ref{k_mode}). This spectral representation of the tidal deformation also allows us to relate the time-derivative expansion of Eq.~(\ref{Q_vs_E_dyn}) to a low-frequency approximation, to explore an $f$-mode approximation to the tidal response (a truncation of the mode sum to a single dominant contribution), and to identify a pragmatic way to extend the time-derivative expansion so that it can successfully capture an approach to resonance. The mode-sum representation of $\tilde{k}_\ell(\omega)$ is not new (see for example, Refs.~\cite{andersson-pnigouras:20, pnigouras-etal:23}), but we choose to derive it {\it ab initio} in order to keep our presentation essentially self-contained. What we do with this representation, especially with regards to the extension of the time-derivative expansion, is new. 

We begin in Sec.~\ref{subsec:moment} by introducing the Lagrangian displacement vector $\bm{\xi}(t,\bm{x})$ and expressing the mass multipole moments ${\cal Q}^{\ell\m}(t)$ in terms of it. The mode-sum representation is introduced in Sec.~\ref{subsec:modesum}, and in Sec.~\ref{subsec:norm} we define the mode norm and overlap integral with the external tidal force. In Sec.~\ref{subsec:frequency} we compute the frequency-domain Love number and arrive at Eq.~(\ref{k_mode}). We define and explore the low-frequency approximation in Sec.~\ref{subsec:fmode}, and in Sec.~\ref{subsec:resum} we propose our pragmatic extension. In Sec.~\ref{subsec:onemode} we reformulate this extension in the form of an effective one-mode approximation for dynamical tides, as seen in Eq.~(\ref{k_onemode}). 

In the course of this discussion we shall turn to the frequency domain, and express the relationship between mass and tidal moments as in Eq.~(\ref{Q_vs_E_freq}), 
\begin{equation}
G \tilde{\cal Q}^{\ell\m}(\omega) = -\frac{2(\ell-2)!}{(2\ell-1)!!} R^{2\ell+1} \tilde{k}_\ell(\omega)\,
\tilde{\cal E}^{\ell\m}(\omega). 
\label{k_omega_def}
\end{equation} 
We use the convention
\begin{equation}
n(t) = \int_{-\infty}^\infty \tilde{n}(\omega) e^{-i\omega t}\, d\omega, \qquad
\tilde{n}(\omega) = \frac{1}{2\pi} \int_{-\infty}^\infty n(t) e^{i\omega t}\, dt
\end{equation}
for the Fourier transform. 

\subsection{Mass moment in terms of Lagrangian displacement}
\label{subsec:moment} 

The perturbation of a fluid is completely determined by the Lagrangian displacement vector $\bm{\xi}(t,\bm{x})$, which takes a fluid element at a position $\bm{x}$ in the unperturbed configuration and places it at position $\bm{x} + \bm{\xi}$ in the perturbed configuration. When the unperturbed configuration is static and spherically symmetric, the vector components in spherical coordinates $(r, \theta^A)$ are given by 
\begin{equation}
\xi_r = \xi^{\ell \m}_r(t,r)\, Y^{\ell \m}(\theta,\phi), \qquad
\xi_A = \xi^{\ell \m}(t,r)\, D_A Y^{\ell \m}(\theta,\phi),
\end{equation}
where $\theta^A := (\theta,\phi)$ and $D_A$ is the covariant-derivative operator compatible with the metric $\Omega_{AB} := \mbox{diag}[1,\sin^2\theta]$ on the unit two-sphere. (When it acts on a scalar, as it does here, the covariant derivative reduces to the partial derivative.) The perturbation of the mass density is given by $\delta \rho = -\nabla_a (\rho \xi^a)$, and this is calculated as
\begin{equation}
\delta \rho = -\frac{1}{r^2} \biggl[ \frac{d}{dr} \bigl( r^2 \rho\, \xi^{\ell\m}_r \bigr)
- \ell(\ell+1) \rho\, \xi^{\ell \m} \biggr] Y^{\ell \m}.
\label{delta_rho} 
\end{equation}
Here we take the perturbation to have a specific value of $\ell$ specified by the applied tidal field; summation over $\m$ is implied. 

We wish to compute the mass moment ${\cal Q}^{\ell\m}(t)$ for a fluid configuration perturbed from a spherical equilibrium. Equation (\ref{resp_potential2}) implies that this is given by
\begin{equation}
{\cal Q}^{\ell\m} = \frac{4\pi \ell!}{(2\ell+1)!!} \int (\rho + \delta \rho )
r^{\ell + 2} \bar{Y}^{\ell\m}\, dr d\Omega, 
\label{Qellm_def}
\end{equation}
where $\rho + \delta \rho$ is the perturbed density, and $d\Omega := \sin\theta\, d\theta d\phi$ is the element of solid angle; an overbar indicates complex conjugation. Noting that the perturbed surface is described by $r = R + \delta R$ with
\begin{equation} 
\delta R = \xi^{\ell \m}_r(r=R)\, Y^{\ell \m}, 
\end{equation}
this is calculated as 
\begin{align}
{\cal Q}^{\ell\m} &= \frac{4\pi \ell!}{(2\ell+1)!!} \int d\Omega \int_0^{R+\delta R}
(\rho + \delta \rho) r^{\ell + 2} \bar{Y}^{\ell \m}\, dr
\nonumber \\
&= \frac{4\pi \ell!}{(2\ell+1)!!} \int d\Omega
\biggl( \int_0^{R+\delta R} \rho r^{\ell + 2} \bar{Y}^{\ell \m}\, dr
+ \int_0^R \delta \rho\, r^{\ell + 2} \bar{Y}^{\ell \m}\, dr \biggr)
\nonumber \\
&= \frac{4\pi \ell!}{(2\ell+1)!!} \int d\Omega
\biggl( \int_R^{R+\delta R} \rho r^{\ell + 2} \bar{Y}^{\ell \m}\, dr 
+ \int_0^R \delta \rho\, r^{\ell + 2} \bar{Y}^{\ell \m}\, dr \biggr);
\end{align}
in the last step we used the fact that the mass moment vanishes for the unperturbed configuration. The first integral returns
\begin{equation}
\rho(r=R) R^{\ell+2} \int \delta R\, \bar{Y}^{\ell\m}\, d\Omega
= \rho(r=R) R^{\ell+2} \xi^{\ell \m}_r(r=R),
\end{equation}
where we used the orthonormality of the spherical harmonics to perform the angular integration. For the second integral we insert Eq.~(\ref{delta_rho}) and get
\begin{equation}
\int d\Omega \int_0^R \delta \rho\, r^{\ell + 2} \bar{Y}^{\ell \m}\, dr
= -\rho(r=R) R^{\ell+2} \xi^{\ell \m}_r(r=R) 
+ \ell \int_0^R \rho \bigl[ r^{\ell+1} \xi^{\ell \m}_r + (\ell+1) r^\ell \xi^{\ell \m} \bigr]\, dr
\end{equation}
after integrating by parts and performing the angular integration. The surface terms cancel out, and the final result is
\begin{equation}
{\cal Q}^{\ell\m} = \frac{4\pi \ell!\, \ell}{(2\ell+1)!!} 
\int_0^R \rho \bigl[ r^{\ell+1} \xi^{\ell \m}_r + (\ell+1) r^\ell \xi^{\ell \m} \bigr]\, dr. 
\label{Qellm_vs_xi}
\end{equation}
The mass moment is now expressed as an integral over components of the Lagrangian displacement vector. 

\subsection{Mode-sum representation}
\label{subsec:modesum} 

The Lagrangian displacement $\bm{\xi}(t,\bm{x})$ is a solution to the perturbation equation (see, for example, Sec.~2.5.3 of Poisson and Will \cite{poisson-will:14})  
\begin{equation}
\partial_{tt} \xi^a + \LL^a_b \xi^b = g^a,
\label{pert_eq} 
\end{equation}
where $\LL^a_b$ is a linear integro-differential operator and $g_a := \nabla_a U^{\rm tidal}$ is the tidal acceleration created by the external objects; the precise identity of $\LL^a_b$ is not required here. A fluid mode $\bm{\zeta}_K(\bm{x})$ is a solution to the time-independent and homogeneous version of the perturbation equation,
\begin{equation}
-\omega_K^2 \zeta^a_K + \LL^a_b \zeta^b_K = 0,
\label{mode_eq} 
\end{equation}
where $K$ is the mode label, and $\omega_K$ is the mode frequency. Modes with different labels are orthogonal, in the sense that
\begin{equation}
\int \rho\, g_{ab} \bar{\zeta}^a_K \zeta^b_{K'}\, dV = N_K\, \delta_{KK'},
\label{ortho}
\end{equation}
where $N_K$ is the mode norm. It is known that the modes form a complete set of basis functions in the Hilbert space defined by Eq.~(\ref{ortho}) \cite{beyer-schmidt:95}. 

We wish to represent $\bm{\xi}(t,\bm{x})$ as a sum over modes. We write
\begin{equation}
\bm{\xi}(t,\bm{x}) = \sum_K q_K(t)\, \bm{\zeta}_K(\bm{x}),
\label{mode_sum} 
\end{equation}
where $q_K(t)$ are mode amplitudes. We similarly write 
\begin{equation}
\bm{g}(t,\bm{x}) = \sum_K g_K(t)\, \bm{\zeta}_K(\bm{x}),
\label{ga_mode} 
\end{equation}
and insert the expansions within Eq.~(\ref{pert_eq}). After using Eq.~(\ref{mode_eq}) and invoking mode completeness, we obtain the oscillator equation
\begin{equation}
\ddot{q}_K + \omega_K^2 q_K = g_K
\label{osc_eq} 
\end{equation}
for each mode amplitude; an overdot indicates differentiation with respect to $t$. The solution to Eq.~(\ref{osc_eq}) is readily expressed as a Fourier transform,
\begin{equation}
q_K(t) = \int_{-\infty}^\infty \tilde{q}_K(\omega) e^{-i\omega t}\, d\omega
\label{q_sol1} 
\end{equation}
with
\begin{equation}
\tilde{q}_K(\omega) = \frac{ \tilde{g}_K(\omega) }{ \omega_K^2 - \omega^2 }, 
\label{q_sol2}
\end{equation}
where $\tilde{g}_K(\omega)$ is the Fourier transform of $g_K(t)$. The mode representation of the perturbation is then obtained by inserting Eqs.~(\ref{q_sol1}) and (\ref{q_sol2}) within Eq.~(\ref{mode_sum}). The mode projection of the tidal acceleration is given by
\begin{equation}
\tilde{g}_K(\omega) = \frac{1}{N_K} \int \rho \tilde{g}_a(\omega) \bar{\zeta}^a_K\, dV;
\label{gK}
\end{equation}
we used the orthogonality relation of Eq.~(\ref{ortho}) to invert Eq.~(\ref{ga_mode}). 

In the current context, in which the unperturbed configuration is static and spherically symmetric, the complete mode label takes the form of $K = n\ell\m$, where $n$ is the overtone label (an integer) and $\ell$, $\m$ are the usual spherical-harmonic integers. A more explicit description of the mode functions is provided by
\begin{equation}
\zeta_r = \zeta_r^{n\ell}(r)\, Y^{\ell\m}(\theta,\phi), \qquad
\zeta_A = \zeta^{n\ell}(r)\, D_A Y^{\ell\m}(\theta,\phi),
\label{zeta_nellm} 
\end{equation}
and the mode frequencies are denoted $\omega_{n\ell}$. By virtue of the spherical symmetry of the unperturbed configuration, the radial functions and frequencies are independent of $\m$.

\subsection{Mode norm and overlap integral}
\label{subsec:norm} 

We proceed with calculations of $N_{n\ell}$ and $g_{n\ell m}$. We shall rely on the orthonormality relations
\begin{subequations}
\label{Y_ortho} 
\begin{align}
\int \bar{Y}^{\ell'\m'} Y^{\ell\m}\, d\Omega &= \delta_{\ell'\ell} \delta_{\m'\m}, \\
\int \Omega^{AB} D_A \bar{Y}^{\ell'\m'} D_B Y^{\ell\m}\, d\Omega &=
\ell(\ell+1)\, \delta_{\ell'\ell} \delta_{\m'\m}
\end{align}
\end{subequations}
for spherical harmonics. 

To calculate the mode norm $N_{n\ell}$ we insert Eq.~(\ref{zeta_nellm}) within Eq.~(\ref{ortho}) and make use of Eq.~(\ref{Y_ortho}) to perform the angular integrations. We arrive at
\begin{equation}
N_{n\ell} = M R^2\, \N_{n\ell}
\label{N_nell1}
\end{equation}
with 
\begin{equation} 
\N_{n\ell} := \frac{1}{MR^2} \int_0^R \rho \Bigl[ r^2 \bigl( \zeta_r^{n\ell} \bigr)^2
+ \ell(\ell+1) \bigl( \zeta^{n\ell} \bigr)^2 \Bigr]\, dr; 
\label{N_nell2}
\end{equation}
the reduced norm $\N_{n\ell}$ is dimensionless.

Next we compute the mode projections $g_{n\ell\m}$ of the tidal acceleration. From Eq.~(\ref{tidal_potential2}) we have that 
\begin{equation}
g_r = -\frac{1}{\ell-1} {\cal E}^{\ell\m} r^{\ell-1} Y^{\ell\m}, \qquad
g_A = -\frac{1}{(\ell-1)\ell} {\cal E}^{\ell\m} r^{\ell} D_A Y^{\ell m}.
\end{equation}
We substitute this within Eq.~(\ref{gK}) and perform the angular integrations. We obtain
\begin{equation}
g_{n\ell\m} = -\frac{1}{\ell-1} R^{\ell-2}\, \frac{\OO_{n\ell}}{\N_{n\ell}}\, {\cal E}^{\ell\m}
\label{g_nell}
\end{equation}
with\footnote{The overlap integral is sometimes defined with an additional factor of $\ell$. See, for example, Eq.~(2.11) of Ref.~\cite{lai:94}.}  
\begin{equation}
\OO_{n\ell} := \frac{1}{MR^\ell} \int_0^R \rho \Bigl[ r^{\ell+1}\, \zeta^{n\ell}_r
+ (\ell+1) r^\ell\, \zeta^{n\ell} \Bigr]\, dr; 
\label{O_nell}
\end{equation}
the reduced overlap integral $\OO_{n\ell}$ is dimensionless.

\subsection{Frequency-domain Love number}
\label{subsec:frequency} 

We return to Eq.~(\ref{Qellm_vs_xi}) for the mass multipole moments. In this we insert Eq.~(\ref{mode_sum}) in the more explicit form
\begin{equation}
\xi_r^{\ell\m}(t,r) = \sum_n q_{n\ell\m}(t)\, \zeta^{\ell\m}_r(r), \qquad
\xi^{\ell\m}(t,r) = \sum_n q_{n\ell\m}(t)\, \zeta^{\ell\m}(r), 
\end{equation}
where $q_{n\ell\m}(t)$ are the mode amplitudes. The radial integral in Eq.~(\ref{Qellm_vs_xi}) is then recognized as $\OO_{n\ell}$, and we obtain
\begin{equation}
{\cal Q}^{\ell\m}(t) = \frac{4\pi \ell!\, \ell}{(2\ell+1)!!} M R^\ell
\sum_n \OO_{n\ell}\, q_{n\ell\m}(t).
\end{equation}
In the final step we transform to the frequency domain, make use of Eq.~(\ref{q_sol2}), and substitute Eq.~(\ref{g_nell}). We obtain
\begin{equation}
\tilde{\cal Q}^{\ell\m}(\omega) = -\frac{4\pi \ell^2 (\ell-2)!}{(2\ell+1)!!} M R^{2\ell-2}\,
\tilde{\cal E}^{\ell\m}(\omega)\, \sum_n \frac{1}{\omega_{n\ell}^2 - \omega^2}
\frac{\OO_{n\ell}^2}{\N_{n\ell}}
\end{equation}
for the frequency-domain multipole moments.

Comparison with Eq.~(\ref{k_omega_def}) allows us to identify the frequency-domain Love number. We find that [Eq.~(\ref{k_mode})] 
\begin{equation}
\tilde{k}_\ell(\omega) = \frac{2\pi \ell^2}{2\ell+1} 
\sum_n \frac{GM/R^3}{\omega_{n\ell}^2 - \omega^2} \frac{\OO_{n\ell}^2}{\N_{n\ell}}, 
\label{k_modesum}
\end{equation}
and have obtained the desired mode-sum representation of the Love number. 

In principle, the mode-sum representation of Eq.~(\ref{k_modesum}) implicates an infinite number of modes. The sum, however, can often be truncated to a finite number of modes without a significant loss of accuracy. At the extreme the sum is truncated to a single term, corresponding to the $n=0$ mode, which is known as the {\it fundamental mode}, or $f$-mode. This mode is characterized by mode functions $\zeta^{0\ell}_r(r)$ and $\zeta^{0\ell}(r)$ with the least number of radial nodes, and comes with the largest overlap integrals. For this reason the $f$-mode is expected to produce the largest contribution to the frequency-domain Love numbers, and the $f$-mode approximation is actually quite good. We shall quantify this statement below. (For an incompressible stellar model with $\rho = \mbox{constant}$, the $f$-mode is the only mode present in the spectrum, and the truncation becomes exact.) 

\subsection{Low-frequency approximation}
\label{subsec:fmode} 

The ratio of time scales $\epsilon$, introduced in Eq.~(\ref{epsilon_def}), will be small whenever $\omega \ll \omega_{n\ell}$ for any mode $n\ell\m$. The time-derivative expansion introduced in Sec.~\ref{sec:newtonian}, therefore, can be viewed as an implementation of a low-frequency approximation. In this regime we have that Eq.~(\ref{k_omega_def}) produces 
\begin{equation}
\label{Q_vs_E_low} 
G{\cal Q}^{\ell\m} (t) = -\frac{2(\ell-2)!}{(2\ell-1)!!} R^{2\ell+1} 
\biggl[ k_\ell\, {\cal E}^{\ell\m} (t) - \ddot{k}_\ell \frac{R^3}{GM}\, \ddot{\cal E}^{\ell\m} (t)
+ \cdots \biggr]
\end{equation}
with
\begin{subequations}
\label{k_ddotk} 
\begin{align} 
k_\ell &:= \tilde{k}_\ell(\omega = 0)
= \frac{2\pi \ell^2}{2\ell+1} 
\sum_n \frac{GM/R^3}{\omega_{n\ell}^2} \frac{\OO_{n\ell}^2}{\N_{n\ell}}, \\
\ddot{k}_\ell &:= \frac{GM}{R^3} \frac{d \tilde{k}_\ell}{d\omega^2} \biggr|_{\omega = 0}
= \frac{2\pi \ell^2}{2\ell+1} 
\sum_n \biggl( \frac{GM/R^3}{\omega_{n\ell}^2} \biggr)^2\,
\frac{\OO_{n\ell}^2}{\N_{n\ell}}. 
\end{align}
\end{subequations}
Equation (\ref{Q_vs_E_low}) is a restatement of Eq.~(\ref{Q_vs_E_dyn}). 

\begin{table}
\caption{\label{tab:ratio} Test of the $f$-mode approximation for a polytropic stellar model with equation of state $p = K \rho^{1+1/n}$. The first column lists the polytropic index $n$, the second the multipole order $\ell$, the third the dimensionless $f$-mode frequency $w_\ell := \omega_{0\ell} (R^3/GM)^{1/2}$, and the fourth the ratio $\rr_\ell := (\ddot{k}_\ell/k_\ell) w_\ell^2$, which is unity when the $f$-mode approximation is exact.}
\begin{ruledtabular}
\begin{tabular}{cccccc}
 $n$ & $\ell$ & $w_\ell$ & $\rr_\ell$ \\  
\hline
1.0 & 2 & 1.226952 & 9.997437$\times 10^{-1}$\\
      & 3 & 1.698253 & 9.992791$\times 10^{-1}$\\
      & 4 & 2.036549 & 9.989909$\times 10^{-1}$\\
      & 5 & 2.310371 & 9.988714$\times 10^{-1}$\\
  & & & \\ 
1.5 & 2 & 1.455807 & 9.978542$\times 10^{-1}$\\
      & 3 & 1.934328 & 9.949372$\times 10^{-1}$\\
      & 4 & 2.258851 & 9.937438$\times 10^{-1}$\\
      & 5 & 2.516478 & 9.936082$\times 10^{-1}$\\
  & & & \\
2.0 & 2 & 1.739606 & 9.891408$\times 10^{-1}$\\
      & 3 & 2.192167 & 9.796660$\times 10^{-1}$\\
      & 4 & 2.483835 & 9.782532$\times 10^{-1}$\\
      & 5 & 2.715511 & 9.797751$\times 10^{-1}$\\
  & & & \\
2.5 & 2 & 2.076292 & 9.580690$\times 10^{-1}$\\
      & 3 & 2.453423 & 9.421301$\times 10^{-1}$\\
      & 4 & 2.697823 & 9.466875$\times 10^{-1}$\\
      & 5 & 2.899605 & 9.544274$\times 10^{-1}$\\
\end{tabular}  
\end{ruledtabular} 
\end{table}

The $f$-mode and low-frequency approximations are independent from one another, but if we choose to combine them, we can simplify Eq.~(\ref{k_ddotk}) to
\begin{subequations}
\begin{align} 
k_\ell &\simeq \frac{2\pi \ell^2}{2\ell+1}
\frac{GM/R^3}{\omega_{0\ell}^2} \frac{\OO_{0\ell}^2}{\N_{0\ell}}, \\
\ddot{k}_\ell &\simeq \frac{2\pi \ell^2}{2\ell+1}
\biggl( \frac{GM/R^3}{\omega_{0\ell}^2} \biggr)^2\, \frac{\OO_{0\ell}^2}{\N_{0\ell}}. 
\end{align}
\end{subequations}
According to this, the static and dynamical Love numbers are related by
\begin{equation}
\ddot{k}_\ell \simeq \frac{GM/R^3}{\omega_{0\ell}^2}\, k_\ell.
\label{k_ddotk_omega} 
\end{equation}
We can test the accuracy of the $f$-mode approximation by computing the mode frequencies $\omega_{0\ell}$ and forming the ratio
\begin{equation}
\rr_\ell := \frac{\ddot{k}_\ell}{k_\ell}\, w^2_\ell, \qquad
w_\ell := \frac{\omega_{0\ell}}{\sqrt{GM/R^3}};
\end{equation}
the approximation will be good when $\rr_\ell$ is sufficiently close to unity. We present the results of this computation in Table~\ref{tab:ratio} for polytropic stellar models with an equation of state $p = K \rho^{1+1/n}$. We see that the $f$-mode approximation is excellent when $n$ is small and the equation of state is relatively stiff: for $n=1$ and for all sampled values of $\ell$, $\rr_\ell$ deviates from unity by no more than $0.1\%$. We see also that the approximation degrades somewhat as $n$ increases and the equation of state becomes softer; for $n=2.5$, $\rr_\ell$ is approximately $5\%$ away from unity. 

\subsection{Beyond the low-frequency approximation}
\label{subsec:resum} 

We now explore how we might go beyond the low-frequency approximation of Eq.~(\ref{Q_vs_E_low}), and attempt to capture more of the oscillator response function, proportional to $(\omega_{n\ell}^2 - \omega^2)^{-1}$. To be concrete we consider a situation in which the tidal field is created by a single companion of mass $M'$ at position $\bm{r}' = r' \bm{n}'$ (with $\bm{n}'$ a unit vector) relative to the reference body. We then have $U^{\rm ext} = GM'/s$ with $s := |\bm{x}-\bm{r}'|$, and Eq.~(\ref{EL_def}) produces
\begin{equation}
{\cal E}_L = -\frac{1}{(\ell-2)!} G M' \partial_L \frac{1}{s} \biggr|_{\bm{x}=\bm{0}}.
\end{equation}
We evaluate the derivatives with the help of Eq.~(1.156) of Poisson and Will \cite{poisson-will:14} and then set $\bm{x} = \bm{0}$. We arrive at
\begin{equation}
{\cal E}_L = -\frac{(2\ell-1)!!}{(\ell-2)!} \frac{GM'}{r^{\prime (\ell+1)}} n'_\stf{L}.
\end{equation}
The equivalent spherical-harmonic representation is
\begin{equation}
{\cal E}^{\ell\m} = -4\pi \frac{(\ell-1) \ell}{2\ell+1} \frac{GM'}{r^{\prime (\ell+1)}}
\bar{Y}^{\ell\m}(\vartheta',\phi'), 
\label{E_companion} 
\end{equation}
where $\theta'$ and $\phi'$ are the polar angles associated with the unit vector $\bm{n}'$. To arrive at Eq.~(\ref{E_companion}) we made use of the definition of Eq.~(\ref{EL_vs_Eell}) and invoked conversion formulae between symmetric-tracefree tensors and spherical harmonics [Eqs.~(1.164) and (1.167) of Poisson and Will].

To be even more concrete (and for the sake of simplicity) we take the companion to move on a circular orbit in the equatorial plane, so that
\begin{equation}
r' = \mbox{constant}, \qquad \theta' = \frac{\pi}{2}, \qquad \phi' = \Omega t, \qquad
\Omega := \sqrt{\frac{G(M+M')}{r^{\prime 3}}}.
\end{equation}
In this case Eq.~(\ref{E_companion}) specializes to
\begin{equation}
{\cal E}^{\ell\m} = -4\pi \frac{(\ell-1) \ell}{2\ell+1} Y^{\ell\m}(\tfrac{\pi}{2},0)\,  
\frac{GM'}{r^{\prime (\ell+1)}} e^{-i \m \Omega t}, 
\label{E_circular} 
\end{equation}
and it follows that
\begin{equation}
\ddot{\cal E}^{\ell\m} = -(\m\Omega)^2\, {\cal E}^{\ell\m}.
\end{equation}
We insert this relation in Eq.~(\ref{Q_vs_E_low}), and obtain
\begin{equation}
G {\cal Q}^{\ell\m}(t) = -\frac{2(\ell-2)!}{(2\ell-1)!!} k_\ell \Gamma_\ell^\m R^{2\ell+1}
{\cal E}^{\ell\m}(t)
\label{Q_vs_E_circ} 
\end{equation}
with
\begin{equation}
\Gamma^\m_{\ell} := 1 + \frac{(\m\Omega)^2}{GM/R^3} \frac{\ddot{k}_\ell}{k_\ell} + \cdots.
\label{Gamma_def}
\end{equation}
Equation (\ref{Q_vs_E_circ}) with $\Gamma^\m_\ell = 1$ is the usual relationship between the mass and tidal multipole moments in the regime of static tides. The additional factor $\Gamma^\m_\ell$ supplies the correction that comes from the dynamical aspects of the tidal interaction.

The expression of Eq.~(\ref{Gamma_def}) is subjected to the low-frequency approximation. In a pragmatic extension of this result, we attempt to capture the high-frequency behavior of the oscillator response function by rewriting $\Gamma^\m_\ell$ as [Eq.~(\ref{Gamma_resum})] 
\begin{equation}
\Gamma^\m_\ell \simeq \biggl[ 1 - \frac{(\m\Omega)^2}{GM/R^3} \frac{\ddot{k}_\ell}{k_\ell} \biggr]^{-1}, 
\label{Gamma_improved} 
\end{equation}
and allowing the expression within brackets to become noticeably smaller than unity. (A zero crossing would signal a gross violation of the low-frequency approximation.) In the context of an inspiral, during which $\Omega$ increases steadily, the $\Gamma^\m_\ell$ of Eq.~(\ref{Gamma_improved}) can grow substantially as the binary approaches merger and $\Omega^2$ becomes comparable to $GM/R^3$. The simple relation of Eq.~(\ref{Gamma_improved}) can therefore capture in an effective way the growing dynamical influence of the tidal interaction in the course of a binary inspiral. And this can be achieved without having to rely on a mode-sum representation of the perturbation.

\subsection{Effective one-mode approximation}
\label{subsec:onemode} 

To proceed with the discussion we insert Eq.~(\ref{Gamma_improved}) within Eq.~(\ref{Q_vs_E_circ}) and take a Fourier transform. We obtain Eq.~(\ref{k_omega_def}), with a frequency-domain Love number given by
\begin{equation}
\tilde{k}_\ell(\omega) = k_\ell \biggl( 1 - \frac{\ddot{k}_\ell/k_\ell}{GM/R^3}\, \omega^2 \biggr)^{-1}.
\label{k_resum1}
\end{equation}
In this specialization to circular orbits, $\tilde{\cal E}^{\ell\m}(\omega)$ is proportional to $\delta(\omega - \m\Omega)$.

We observe that Eq.~(\ref{k_resum1}) bears a formal resemblance to Eq.~(\ref{k_modesum}) when the mode sum is truncated to a single term. We may therefore think of Eq.~(\ref{k_resum1}) as the embodiment of an {\it effective one-mode approximation}. We make this manifest by introducing an effective mode frequency $\omega_{*\ell}$ defined by 
\begin{equation}
\omega^2_{*\ell} := \frac{GM}{R^3} \frac{k_\ell}{\ddot{k}_\ell},
\label{omega_star} 
\end{equation}
and rewriting Eq.~(\ref{k_resum1}) as
\begin{equation}
\tilde{k}_\ell(\omega) = \frac{\tilde{k}_\ell(0)}{1 - \omega^2/\omega^2_{*\ell}},
\label{k_resum2} 
\end{equation}
which is now fully equivalent to the one-mode version of Eq.~(\ref{k_modesum}); we recall that $k_\ell := \tilde{k}_\ell(\omega=0)$. The notation $\omega_{*\ell}$ for the effective mode frequency reminds us that conceptually, this quantity is not to be associated with any of the star's normal modes of vibration; it is a mode-less creature of our extended low-frequency approximation. The notation is meant to suggest that the assignment of Eq.~(\ref{Gamma_improved}) can be formally related to a one-mode truncation of Eq.~(\ref{k_modesum}).

But we actually have more than this. Our test of the $f$-mode approximation established that the effective mode frequency $\omega_{*\ell}$ is numerically very close to the star's $f$-mode frequency. This reveals that the effective one-mode approximation is very closely related to the $f$-mode approximation, and that the accuracy of Eq.~(\ref{Gamma_improved}) can be equated with the accuracy of the $f$-mode approximation. Because the latter was shown earlier to be quite accurate, we have compelling evidence that Eq.~(\ref{Gamma_improved}) should be just as accurate. 

Our conclusion is that the pragmatic extension of the low-frequency approximation given by Eq.~(\ref{Gamma_improved}) provides a description of dynamical tides that compares very well in accuracy with a mode sum truncated to include just the $f$-mode. And as we have seen, this level of accuracy is perfectly adequate.    

\section{Static and dynamic Love numbers in general relativity}
\label{sec:GR}

In this section we compute the static and dynamic Love numbers of a polytropic star in full general relativity. We begin in Sec.~\ref{subsec:exterior} with a review of the exterior metric of a tidally deformed material body, as constructed in Ref.~\cite{poisson:21a}; the metric provides a precise, relativistic definition for the Love numbers. Next we introduce the interior variables (Sec.~\ref{subsec:interior}), derive the relevant perturbation equations (Sec.~\ref{subsec:structure}), and specialize them to polytropic stellar models (Sec.~\ref{subsec:polytropeGR}). In Sec.~\ref{subsec:LoveGR} we describe how the matching of the interior and exterior metrics at the stellar surface produces the Love numbers.

\subsection{Exterior metric}
\label{subsec:exterior} 

We consider a nonrotating, tidally deformed, material body of mass $M$ and radius $R$. We describe its gravitational field in full general relativity, in terms of a metric tensor $g_{\alpha\beta}$. The tidal environment is again characterized by tidal moments ${\cal E}^{\ell\m}(t)$ that vary slowly with time. The nonvanishing components of the metric are expressed as
\begin{equation}
g_{tt} = -f + p_{tt}\, \qquad
g_{tr} = p_{tr}\, \qquad
g_{rr} = f^{-1} + p_{rr}\, \qquad
g_{AB} = r^2 \Omega_{AB} (1 + q),
\label{tidal_metric} 
\end{equation}
where $f := 1-2M/r$, $\theta^A = (\theta,\phi)$, $\Omega_{AB} = \mbox{diag}[1,\sin^2\theta]$, and where
\begin{subequations}
\label{metric_pert}
\begin{align}
p_{tt} &= -\frac{2}{(\ell-1)\ell} f^2 r^\ell \Bigl( e_{tt}\, {\cal E}^{\ell\m}
+ \ddot{e}_{tt}\, M^2 \ddot{\cal E}^{\ell\m} \Bigr) Y^{\ell\m}, \\
p_{tr} &= -\frac{4}{(\ell-1)\ell(\ell+1)} f^{-1} r^{\ell+1} \dot{e}_{tr}\,
\dot{\cal E}^{\ell\m}\, Y^{\ell\m}, \\
p_{rr} &= -\frac{2}{(\ell-1)\ell} r^\ell \Bigl( e_{rr}\, {\cal E}^{\ell\m}
+ \ddot{e}_{rr}\, M^2 \ddot{\cal E}^{\ell\m} \Bigr) Y^{\ell\m}, \\
q &= -\frac{2}{(\ell-1)\ell} r^\ell \Bigl( e\, {\cal E}^{\ell\m}
+ \ddot{e}\, M^2 \ddot{\cal E}^{\ell\m} \Bigr) Y^{\ell\m}
\end{align}
\end{subequations}
are the components of the metric perturbation, presented in the Regge-Wheeler gauge \cite{regge-wheeler:57} (see also Ref.~\cite{martel-poisson:05} for a formulation of the theory of gravitational perturbations of a Schwarzschild spacetime). An overdot on the tidal moment ${\cal E}^{\ell\m}$ continues to indicate differentiation with respect to $t$, and $Y^{\ell\m}(\theta,\phi)$ continues to denote spherical harmonics.  

The radial functions associated with the static perturbation are given by
\begin{subequations}
\label{static_functions}
\begin{align}
e_{tt} &= e_{rr} = A_\ell + 2 K_\ell (M/r)^{2\ell+1}\, B_\ell, \\
e &= C_\ell + 2 K_\ell (M/r)^{2\ell+1}\, D_\ell,
\end{align}
\end{subequations}
with functions $A_\ell(r)$, $B_\ell(r)$, $C_\ell(r)$, and $D_\ell(r)$ defined by Eq.~(5.8) of Ref.~\cite{poisson:21a} in terms of hypergeometric functions. Explicit expressions for $\ell = \{2,3,4,5\}$ appear in Appendix C of this reference. Each one of these functions behaves as $1 + O(M/r)$ when expanded in powers of $M/r$. The quantity $K_\ell$ is a rescaled Love number, related by 
\begin{equation}
K_\ell = k_\ell (R/M)^{2\ell+1}
\end{equation}
to the scalefree Love number $k_\ell$. 

The radial functions associated with the dynamic perturbation are expressed as
\begin{subequations}
\label{dynamic_functions}
\begin{align}
\ddot{e}_{tt} &= \ddot{e}_{rr} = \ddot{T}_\ell\, A_\ell - 2 \ddot{K}_\ell (M/r)^{2\ell+1}\, B_\ell
+ \gotha_\ell + K_\ell\, \gothb_\ell, \\
\ddot{e} &= \ddot{T}_\ell\, C_\ell - 2 \ddot{K}_\ell (M/r)^{2\ell+1}\, D_\ell
+ \gothc_\ell + K_\ell\, \gothd_\ell,
\end{align}
\end{subequations}
with a new set of functions $\gotha_\ell(r)$, $\gothb_\ell(r)$, $\gothc_\ell(r)$, and $\gothd_\ell(r)$ that are listed explicitly for $\ell = \{2,3,4,5\}$ in Appendix F of Ref.~\cite{poisson:21a}. The quantity $\ddot{K}_\ell$ is a rescaled Love number, related to the scalefree Love number $\ddot{k}_\ell$ by
\begin{equation}
\ddot{K}_\ell = \ddot{k}_\ell (R/M)^{2\ell+4}.
\end{equation}
The remaining function $\dot{e}_{tr}$ is defined in terms of hypergeometric functions in Eq.~(5.47) of Ref.~\cite{poisson:21a}; explicit expressions for $\ell = \{2,3,4,5\}$ appear in Appendix E of this reference. We note that the dynamic Love number is defined here with a minus sign relative to the original definition of Ref.~\cite{poisson:21a}; we introduced the same minus sign in the Newtonian discussion of Sec.~\ref{sec:newtonian}. The current definition ensures that the numbers turn out to be positive. 

The meaning of the constant $\ddot{T}_\ell$ is explained in Sec.~V K of Ref.~\cite{poisson:21a}. It is arbitrary, and the freedom to choose its value is associated with the freedom to redefine the tidal moments according to
\begin{equation}
{\cal E}^{\ell\m} \to  {\cal E}^{\ell\m} + \ddot{\lambda}_\ell M^2 \ddot{\cal E}^{\ell\m},
\label{moment_redefinition1} 
\end{equation}
where $\ddot{\lambda}_\ell$ is another arbitrary constant. It is easy to show that the impact of this transformation is to produce the changes
\begin{equation}
\ddot{T}_\ell \to \ddot{T}_\ell + \ddot{\lambda}_\ell, \qquad 
\ddot{K}_\ell \to \ddot{K}_\ell + \ddot{\lambda}_\ell K_\ell
\label{moment_redefinition2} 
\end{equation}
in the constants that appear in the perturbed metric. In the Newtonian limit $M/r \ll 1$ the metric perturbation of Eq.~(\ref{metric_pert}) should reduce to the Newtonian potential of Eq.~(\ref{Uext_dyn}), and this is achieved when we make the choice $\ddot{T}_\ell = 0$. The freedom to redefine the tidal moments can therefore be exercised to set $\ddot{T}_\ell = 0$, and once this is done, the dynamic Love numbers become invariant. We do not make this choice at the outset, because to keep $\ddot{T}_\ell$ arbitrary in the exterior metric is helpful for the computations to be presented below; we shall explain why in Sec.~\ref{subsec:polytropeGR}. 

The metric of Eq.~(\ref{tidal_metric}) is a special case of the metric constructed in Ref.~\cite{poisson:21a}. The more complete version includes additional terms that are proportional to the first derivative of the tidal moments. These terms break the time-reversal invariance of the metric, and are required when the physics of the tidally deformed body includes dissipation (as in the case of a viscous fluid). In our case the body is modelled as a perfect fluid, the physics is time-reversal invariant, and the additional terms are appropriately excluded.

\subsection{Interior metric and fluid variables}
\label{subsec:interior} 

We write the metric inside the body as 
\begin{equation}
g_{tt} = -e^{2\psi} + p_{tt}\, \qquad
g_{tr} = p_{tr}\, \qquad
g_{rr} = f^{-1} + p_{rr}\, \qquad
g_{AB} = r^2 \Omega_{AB} (1 + q),
\label{tidal_metric_inside} 
\end{equation}
where $\psi = \psi(r)$ and $f := 1-2m(r)/r$, with $m(r)$ standing for the mass inside a sphere of radius $r$. The metric perturbation is written as
\begin{subequations}
\label{pert_inside}
\begin{align}
p_{tt} &= -\frac{2}{(\ell-1)\ell} e^{2\psi} f r^\ell \Bigl\{ a_\ell\, {\cal E}^{\ell\m}
+ e^{-2\psi} f \ddot{a}_\ell\, r_1^2 \ddot{\cal E}^{\ell\m} \Bigr\} Y^{\ell\m}, \\
p_{tr} &= -\frac{4}{(\ell-1)\ell(\ell+1)} f^{-1} r^{\ell+1}\, \dot{b}_\ell\,
\dot{\cal E}^{\ell\m}\, Y^{\ell\m}, \\
p_{rr} &= -\frac{2}{(\ell-1)\ell} r^\ell \Bigl\{ a_\ell\, {\cal E}^{\ell\m}
+ e^{-2\psi} f \ddot{a}_\ell\, r_1^2 \ddot{\cal E}^{\ell\m} \Bigr\} Y^{\ell\m}, \\
q &= -\frac{2}{(\ell-1)\ell} r^\ell \Bigl\{ \bigl[ a_\ell + (r/r_1)^2 c_\ell \bigr]\, {\cal E}^{\ell\m}
+ e^{-2\psi} f \bigl[ \ddot{a}_\ell + (r/r_1)^2 \ddot{c}_\ell \bigr] r_1^2
\ddot{\cal E}^{\ell\m} \Bigr\} Y^{\ell\m},
\end{align}
\end{subequations}
where the coefficients $a_\ell$, $c_\ell$, $\dot{b}_\ell$, $\ddot{a}_\ell$, and $\ddot{c}_\ell$ are functions of $r$ only. The metric perturbation is presented in Regge-Wheeler gauge, as it was for the exterior metric. The form displayed in Eq.~(\ref{pert_inside}) anticipates an outcome of imposing the Einstein field equations, that $e^{-2\psi} p_{tt}$ must be equal to $f p_{rr}$. We have peppered the expressions with numerical factors, powers of $r$, and factors of $e^{2\psi}$ and $f$ in order to (i) ensure that all radial functions approach a constant when $r \to 0$, (ii) eliminate all factors of $e^{2\psi}$ in the perturbation equations, and (iii) facilitate the matching of the interior perturbation with the exterior perturbation at $r=R$. The split of $q$ in terms of radial functions $a_\ell$ and $c_\ell$ (as well as $\ddot{a}_\ell$ and $\ddot{c}_\ell$) reflects the fact that the functions that appear in $p_{tt}$ and $q$ share the same limit when $r \to 0$, but differ at order $r^2$. Finally, we have inserted a length scale $r_1$ within the metric perturbation, to compensate dimensionally for the time derivatives of the tidal moments; this scale is arbitrary, and it will be chosen at a later stage.

The stellar matter is modelled as a perfect fluid, and it possesses an energy-momentum tensor
\begin{equation}
T^{\alpha\beta} = (\mu + \delta\mu) u^\alpha u^\beta
+ (p + \delta p) \bigl( g^{\alpha\beta} + u^\alpha u^\beta \bigr),
\end{equation}
where $\mu + \delta\mu$ is the perturbed energy density, $p + \delta p$ is the perturbed pressure, $u^\alpha$ is the perturbed velocity vector, and $g^{\alpha\beta}$ is the perturbed inverse metric. We write the components of the velocity vector as
\begin{subequations}
\label{velocity_vector}
\begin{align} 
u^t &= e^{-\psi} - \frac{1}{(\ell-1)\ell} e^{-\psi} f r^\ell \Bigl\{ a_\ell\, {\cal E}^{\ell\m}
+ e^{-2\psi} f\, \ddot{a}_\ell\, r_1^2 \ddot{\cal E}^{\ell\m} \Bigr\} Y^{\ell\m}, \\ 
u^r &= \frac{2}{(\ell-1)\ell} e^{-\psi} r^{\ell-1}\, \dot{u}_\ell\, r_1^2 \dot{\cal E}^{\ell\m}\, Y^{\ell\m}, \\
u^A &= \frac{2}{(\ell-1)\ell} e^{-\psi} r^{\ell-2}\, \dot{v}_\ell\, r_1^2 \dot{\cal E}^{\ell\m}\,
\Omega^{AB} D_B Y^{\ell\m},
\end{align}
\end{subequations}
where $\dot{u}_\ell$ and $\dot{v}_\ell$ are new radial functions, $\Omega^{AB} := \mbox{diag}[1,1/\sin^2\theta]$ is the matrix inverse to $\Omega_{AB}$, and $D_A$ is the covariant-derivative operator compatible with $\Omega_{AB}$. The expression for $u^t$ follows from the normalization condition $g_{\alpha\beta} u^\alpha u^\beta = -1$. The factor of $r^{\ell-1}$ in $u^r$, and the factor of $r^{\ell-2}$ in $u^A$, are inherited from the Newtonian expressions considered previously in Sec.~\ref{sec:newtonian}. 

We take the fluid to satisfy a barotropic equation of state, which we write in the form $\mu = \mu(p)$. The angular components of the conservation equation $\nabla_\beta T^{\alpha\beta} = 0$ (with a covariant derivative compatible with the perturbed metric) reveal that the pressure perturbation can be written as
\begin{equation}
\delta p = -\frac{1}{(\ell-1)\ell} (\mu+p) f r^\ell \Bigl\{ a_\ell\, {\cal E}^{\ell\m}
+ e^{-2\psi} \bigl( f \ddot{a}_\ell + 2 f^{-1} \dot{v}_\ell \bigr) r_1^2 \ddot{\cal E}^{\ell\m} \Bigr\} Y^{\ell\m}, 
\end{equation}
in terms of the metric and velocity perturbation. The equation of state then delivers $\delta \mu = (d\mu/dp)\, \delta p$.

The complete listing of perturbation variables consists of $\{ a_\ell, c_\ell \}$ for a static perturbation, $\{ \dot{u}_\ell, \dot{v}_\ell, \dot{b}_\ell \}$ for the perturbation constructed from the first derivative of the tidal moments, and $\{ \ddot{a}_\ell, \ddot{c}_\ell \}$ for the perturbation associated with the second derivatives. All variables are dimensionless. We shall see below that $\dot{b}_\ell$ is algebraically related to other variables, and can therefore be eliminated from the list of independent radial functions. The remaining number is then six.

\subsection{Structure and perturbation equations}
\label{subsec:structure} 

The Einstein field equations for an unperturbed metric and fluid return the structure equations
\begin{subequations}
\label{structure}
\begin{align}
\frac{dm}{dr} &= 4\pi r^2 \mu, \\ 
\frac{d\psi}{dr} &= \frac{m+4\pi r^3 p}{r^2 f}, \\ 
\frac{dp}{dr} &= -(\mu+p) \frac{d\psi}{dr}.  
\end{align}
\end{subequations} 
These can be integrated as soon as an equation of state $\mu = \mu(p)$ is supplied. We note that the equation for $\psi$ will be used freely in subsequent developments, but that it shall never be integrated explicitly.

Turning next to the equations governing $a_\ell$ and $c_\ell$, we observe that an independent set of equations is provided by the $rr$ and $rA$ components of the Einstein field equations; the remaining components deliver redundant information. We obtain
\begin{subequations}
\label{0time}
\begin{align}
f (m+4\pi r^3 p) \frac{da_\ell}{dr} &= \Bigl[ 4\pi r^2(\mu+p)(1 + 8\pi r^2 p)
+ 2\ell m^2/r^2 + 8 \pi (\ell-4) r p m 
\nonumber \\ & \quad \mbox{} 
+ (\ell-2)(\ell+2) m/r - 4\pi\ell r^2 p - 64\pi^2 r^4 p^2 \Bigr] a_\ell
+ \frac{1}{2} (\ell-1)(\ell+2) (r/r_1)^2\, c_\ell, \\
f (m+4\pi r^3 p) \frac{dc_\ell}{dr} &= -2(r_1/r)^2 \Bigl[ 8\pi (\mu+p)(2\pi r^4 p - m^2 + rm)
+ 4m^3/r^3 + (\ell-2)(\ell+3) m^2/r^2
\nonumber \\ & \quad \mbox{} 
- 16\pi r p m - 32\pi^2 r^4 p^2 \Bigr] a_\ell
+ (\ell+2) \Bigl[ 2m^2/r^2 - (\ell-8\pi r^2 p) m/r - 4\pi r^2 p \bigr] c_\ell.
\end{align}
\end{subequations}
The differential equations are singular at $r=0$. A local analysis reveals that regular solutions for $a_\ell$ and $c_\ell$ tend to constants when $r \to 0$; the constants are related to each other, so that only one is independent. The equations are regular as $r \to R$, and there is no need to impose specific boundary conditions at the surface.

Moving on to the equations that govern $\dot{u}_\ell$ and $\dot{v}_\ell$, we have that the $t$ and $r$ components of $\nabla_\beta T^{\alpha\beta} = 0$ deliver
\begin{subequations}
\label{1time1} 
\begin{align} 
r \frac{d\dot{u}_\ell}{dr} &= f^{-1} \Bigl[ (m/r + 4\pi r^2 p) d\mu/dp
- (\ell+1) + (2\ell+3)m/r - 4\pi r^2\mu \Bigr] \dot{u}_\ell + \ell(\ell+1) \dot{v}_\ell
\nonumber \\ & \quad \mbox{}
+ \frac{1}{2} (r/r_1)^2 \Bigl[ 3 - 2m/r + f d\mu/dp \Bigr] a_\ell
+ (r/r_1)^4\, c_\ell, \\ 
r f \frac{d\dot{v}_\ell}{dr} &= \dot{u}_\ell - \Bigl[ \ell - 2(\ell+1) m/r - 8\pi r^2 p \Bigr] \dot{v}_\ell
  - \frac{2}{\ell+1} (r/r_1)^2\, \dot{b}_\ell, 
\end{align}
\end{subequations}
while the $tr$ component of the Einstein field equations produces
\begin{align}
\dot{b}_\ell &= \frac{r}{\ell(m + 4\pi r^3 p)} \biggl\{
\Bigl[ 4\pi r^2 (\mu + p) (1 - 4m/r + 4m^2/r^2) - 12m^3/r^3 - (2\ell^2+2\ell-13) m^2/r^2
- 16\pi p m^2
\nonumber \\ & \quad \mbox{}
+ (\ell^2+\ell-4) m/r - 16\pi^2 r^4 p^2 \Bigr] a_\ell
- \frac{1}{2} (r/r_1)^2 \Bigl[ 6 m^2/r^2 + 32\pi r p m + 2(\ell^2+\ell-3) m/r
+ 32\pi^2 r^4 p^2
\nonumber \\ & \quad \mbox{}
- 8\pi r^2 p - (\ell-1)(\ell+2) \Bigr] c_\ell
+ 8 \pi r_1^2\, (\mu+p) (m/r + 4\pi r^2 p)\, \dot{u}_\ell \biggr\},
\label{1time2} 
\end{align} 
an algebraic equation for $\dot{b}_\ell$. The equations reveal that $\dot{u}_\ell$ and $\dot{v}_\ell$ both tend to a constant when $r \to 0$, and that these are related by $\dot{u}_\ell(r=0) = \ell\, \dot{v}_\ell(r=0)$. The presence of $d\mu/dp$ in the differential equation for $\dot{u}_\ell$ implies that the system of equations is singular at $r=R$; the solution will be well behaved if we impose 
\begin{equation}
(M/R)\, \dot{u}_\ell(r=R) = -\frac{1}{2} (R/r_1)^2 (1-2M/R)^2\, a_\ell(r=R)
\label{1time3}
\end{equation}
as a boundary condition. 

Finally, the equations that govern $\ddot{a}_\ell$ and $\ddot{c}_\ell$ come from the $rr$ and $rA$ components of the field equations, with the remaining equations providing redundant information. We get
\begin{subequations}
\label{2time}
\begin{align}
f (m+4\pi r^3 p) \frac{d\ddot{a}_\ell}{dr} &= \Bigl[ 4\pi r^2 (\mu+p) (1 + 2m/r + 16\pi r^2 p)
+ 2\ell m^2/r^2 + (\ell-2)(\ell+2) m/r
\nonumber \\ & \quad \mbox{}
+ 8\pi (\ell-4) rpm - 64\pi^2 r^4 p^2 - 4\pi \ell r^2 p \Bigr] \ddot{a}_\ell
+ \frac{1}{2} (\ell-1)(\ell+2) (r/r_1)^2\, \ddot{c}_\ell
\nonumber \\ & \quad \mbox{}
- \frac{2}{(\ell+1) f^2} (r/r_1)^2 \Bigl[ 1 - 3m/r - 4\pi r^2 p \Bigr] \dot{b}_\ell
+ 8\pi f^{-1} r^2 (\mu+p)\, \dot{v}_\ell
\nonumber \\ & \quad \mbox{}
+ (r/r_1)^2 f^{-1} \Bigl[ a_\ell + (r/r_1)^2\, c_\ell \Bigr], \\
f (m+4\pi r^3 p) \frac{d\ddot{c}_\ell}{dr} &= (r_1/r)^2 \Bigl[ 16\pi r^2 (\mu+p)(m^2/r^2 - m/r - 2\pi r^2 p)
- 8m^3/r^3 - 2(\ell-2)(\ell+3) m^2/r^2
\nonumber \\ & \quad \mbox{}
+ 32\pi rpm + 64\pi^2 r^4 p^2 \Bigr] \ddot{a}_\ell
+ \Bigl[ 8\pi r^2 (\mu+p) (m/r + 4\pi r^2 p) + (\ell+2) ( 2m^2/r^2 
\nonumber \\ & \quad \mbox{}
- \ell m/r + 8\pi rpm - 4\pi r^2 p ) \Bigr] \ddot{c}_\ell
- \frac{2}{(\ell+1) f^2} \Bigl[ 4m^2/r^2 - m/r + 4\pi r^2 p \Bigr] \dot{b}_\ell
\nonumber \\ & \quad \mbox{}
- 16\pi r_1^2 f^{-1} (\mu+p) (m/r)\, \dot{v}_\ell
- 2 f^{-1} (m/r) \Bigl[ a_\ell + (r/r_1)^2 c_\ell \Bigr].
\end{align}
\end{subequations}
We have that $\ddot{a}_\ell$ and $\ddot{c}_\ell$ both tend to a constant as $r \to 0$, with only one of them independent. The equations are regular at $r=R$, and there is no need to impose a boundary condition there.

The equations listed in this subsection form a complete set for all structure and perturbation variables. Once a solution is at hand, matching the metric perturbation at $r=R$ with the exterior solution of the preceding subsection determines the unknown quantities associated with the exterior metric, namely the Love numbers $k_\ell$ and $\ddot{k}_\ell$, as well as the additional constant $\ddot{T}_\ell$. 

\subsection{Implementation for a polytrope}
\label{subsec:polytropeGR} 

At this stage we must specify an equation of state, and once again we adopt the simple polytropic form $p = K \rho^{1+1/n}$, where $\rho$ is the fluid's particle mass density (particle number density times the average rest-mass of the constituent particle), and where $K$ and $n$ are constants. The first law of thermodynamics, $d(\epsilon/\rho) + p d(1/\rho) = 0$, implies that the internal energy density is given by $\epsilon = n p$ for this equation of state. The total energy density is then $\mu = \rho + \epsilon$.

We introduce a dimensionless substitute $\vartheta$ for the density, by writing
\begin{equation}
\rho = \rho_c\, \vartheta^n, \qquad
p = b\rho_c\, \vartheta^{n+1}, \qquad
\mu = \rho_c \vartheta^n(1 + n b \vartheta),
\end{equation}
where $\rho_c := \rho(r=0)$ is the central density, and $b := p_c/\rho_c := p(r=0)/\rho(r=0)$ is the ratio of central pressure to central density. In terms of the new variable we have that
\begin{equation}
\frac{d\mu}{dp} = n \biggl[ 1 + \frac{1}{(n+1)b \vartheta} \biggr],
\end{equation}
and we see that this diverges when $\vartheta \to 0$ (when the density vanishes). We also introduce dimensionless substitutes $\chi$ and $\zeta$ for the mass function and radial coordinate, respectively, by writing
\begin{equation}
m =\frac{4\pi}{3} \rho_c (1+nb) r^3\, \chi, \qquad
r^2 = \frac{3}{2\pi} \frac{(n+1) b}{(1+nb) \rho_c}\, \zeta. 
\end{equation}
From Eq.~(\ref{structure}) we get that the structure equations become
\begin{subequations}
\label{structure_dimless}
\begin{align}
\frac{d\vartheta}{d\zeta} &= -\frac{1}{f} \bigl[1 + (n+1)b \vartheta \bigr]
\biggl( \chi + \frac{3b}{1+nb} \vartheta^{n+1}\biggr), \\
\frac{d\chi}{d\zeta} &= -\frac{3}{2\zeta} \biggl( \chi - \frac{1+nb\vartheta}{1+nb} \vartheta^n \biggr), 
\end{align} 
\end{subequations}
where $f := 1-2m/r = 1 - 4(n+1)b \zeta \chi$. It is very convenient to use $\vartheta$ as the independent variable instead of $\zeta$ \cite{lindblom:92}. In this formulation, Eqs.~(\ref{structure_dimless}) are integrated from the centre at $\vartheta = 1$ with the starting values $\zeta(\vartheta = 1) = 0$ and $\chi(\vartheta = 1) = 1$, up to the surface at $\vartheta = 0$, where we obtain $\zeta_s := \zeta(\vartheta = 0)$ and $\chi_s := \chi(\vartheta = 0)$. The star's global quantities $M := m(r=R)$ and $R$ are obtained from these, and the star's compactness is 
\begin{equation}
M/R = 2(n+1) b\, \zeta_s \chi_s. 
\end{equation}
An equilibrium sequence is obtained by integrating the structure equations for a range of central densities $\rho_c$. In practice it is more convenient to parametrize the sequence with $b$, which is in a one-to-one relationship with $\rho_c$ for the selected equation of state. The sequence ends at the configuration of maximum mass; beyond this point the equilibria are dynamically unstable to radial perturbations. 

It is a straightforward matter to rewrite the perturbation equations (\ref{0time}), (\ref{1time1}), (\ref{1time2}), and (\ref{2time}) in terms of the dimensionless variables $\theta$, $\zeta$, and $\chi$; we shall not provide these details here. As a convenient choice of length scale $r_1$ we set  
\begin{equation}
r_1^2 = \frac{3}{2\pi} \frac{n+1}{(1+nb) \rho_c}. 
\end{equation}
This differs by a factor of $1/b$ from the scaling factor previously introduced in the definition of $\zeta$. This choice is motivated by the desire to keep the equations numerically well behaved when $b$ becomes small; this limit takes us to a Newtonian body.

We observed previously that each perturbation variable approaches a constant when $r \to 0$, and that there are relations among these constants. For the polytropic equation of state we find that these are given by
\begin{subequations}
\label{BD_centre}
\begin{align}
c_\ell(r=0) &= -\frac{4(n+1)}{(\ell+2)(1 + nb)} \Bigl\{ \ell + 1 + \bigl[ (\ell+1)n - 3 \bigr] b \Bigr\}\, a_\ell(r=0),
\label{BD_centre_a} \\
\dot{u}_\ell(r=0) &= \ell\, \dot{v}_\ell(r=0),
\label{BD_centre_b} \\
\ddot{c}_\ell(r=0) &= -\frac{4}{(\ell+1)(\ell+2)(1 + nb)} \Bigl\{ \bigl[ (\ell+1)^2 n^2
+ (\ell^2-\ell-2) n - 3(\ell+1) \bigr] b + (\ell+1)^2(n + 1) \Bigr\} \ddot{a}_\ell(r=0)
\nonumber \\ & \quad \mbox{} 
- \frac{24(n+1)\bigl[(n+1)b + 1\bigr]}{\ell(\ell+1)(\ell+2)(1+nb)}\, \dot{u}_\ell(r=0) 
- \frac{2}{(\ell+1)(\ell+2)}\, a_\ell(r=0)
\label{BD_centre_c} 
\end{align}
\end{subequations}
and
\begin{equation}
\dot{u}_\ell(r=R) = -\biggl[ 4(n+1) b^2\, \zeta_s^2 \chi_s - 2b\, \zeta_s
+ \frac{1}{4(n+1)\, \chi_s} \biggr] a_\ell(r=R).
\label{BD_surface}
\end{equation}
Equation (\ref{BD_surface}) is a restatement of Eq.~(\ref{1time3}). 

The integration of Eqs.~(\ref{0time}) for $a_\ell$ and $c_\ell$ requires the specification of a single arbitrary constant, $a_\ell(r=0)$; $c_\ell(r=0)$ is then determined by Eq.~(\ref{BD_centre_a}). The selection of $a_\ell(r=0)$ provides an overall normalization to the solution, and this choice of normalization will be propagated through the remaining perturbation variables. We set $a_\ell(r=0) = 1$.

Moving on to the integration of Eqs.~(\ref{1time1}) for $\dot{u}_\ell$ and $\dot{v}_\ell$, we recall that the system is singular at both $r=0$ and $r=R$, and we must now impose boundary conditions at both ends. These are given by Eqs.~(\ref{BD_centre_b}) and Eq.~(\ref{BD_surface}). The boundary conditions ensure that the solution is unique; there is no freedom of choice with these perturbation variables.

The system of Eqs.~(\ref{2time}) for $\ddot{a}_\ell$ and $\ddot{c}_\ell$ is singular at $r=0$, and we have a single boundary condition there, given by Eq.~(\ref{BD_centre_c}). In this equation, $a_\ell(r=0) = 1$, $\dot{u}_\ell(r=0)$ is known from the preceding integrations, and $\ddot{a}_\ell(r=0)$ is arbitrary. The freedom to choose $\ddot{a}_\ell(r=0)$ corresponds to the freedom to add to $\ddot{a}_\ell$ and $\ddot{c}_\ell$ a solution to the homogeneous version of Eqs.~(\ref{2time}), obtained by removing all source terms proportional to $a_\ell$, $c_\ell$, $\dot{u}_\ell$, and $\dot{v}_\ell$. And in turn, this corresponds to the freedom to redefine the tidal moments according to Eq.~(\ref{moment_redefinition1}). Now, we recall that the exterior solution to the perturbation equations involved a number $\ddot{T}_\ell$ that was also associated with the freedom to redefine the tidal moments. The choice of $\ddot{a}_\ell(r=0)$, therefore, will be tied to a determination of $\ddot{T}_\ell$ when the interior perturbation is matched to the exterior perturbation. Because $\ddot{T}_\ell$ is entirely arbitrary, we have the freedom to set $\ddot{a}_\ell(r= 0)$ arbitrarily, and we shall impose $\ddot{a}_\ell(r= 0) = 1$. In the final step of the computation, we will implement the transformation of Eqs.~(\ref{moment_redefinition1}) and (\ref{moment_redefinition2}) to set the final value of $\ddot{T}_\ell$ to zero, and obtain a final value for $\ddot{K}_\ell$ that is invariant under a redefinition of the tidal moments.

Additional details regarding the numerical integration of the perturbation equations are given in Appendix~\ref{sec:numerical}. We carried out two independent integrations, one exploiting collocation methods based on an expansion of all variables in Chebyshev polynomials, the other using finite-difference methods. Agreement between these results gives us confidence that the computations are accurate. 

\subsection{Love numbers}
\label{subsec:LoveGR} 

Integration of the perturbation equations returns surface values $a_s := a_\ell(r=R)$, $c_s := c_\ell(r=R)$, and so on, for all the perturbation variables. Matching the interior solution with the exterior solution at $r=R$ allows us to determine the Love numbers $k_\ell$ and $\ddot{k}_\ell$, as well as the arbitrary constant $\ddot{t}_\ell := (M/R)^3 \ddot{T}_\ell$.

\begin{figure}
\includegraphics[width=0.49\linewidth]{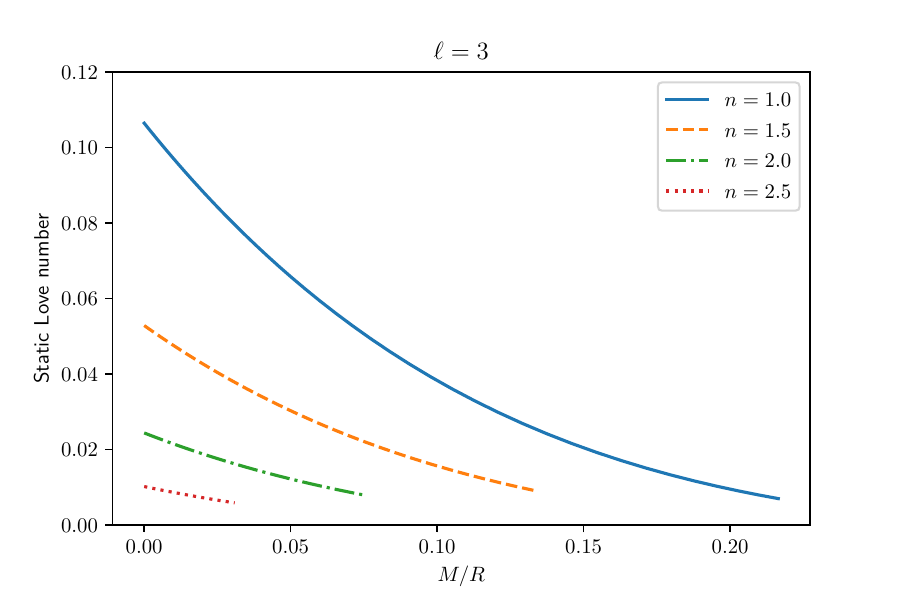}
\includegraphics[width=0.49\linewidth]{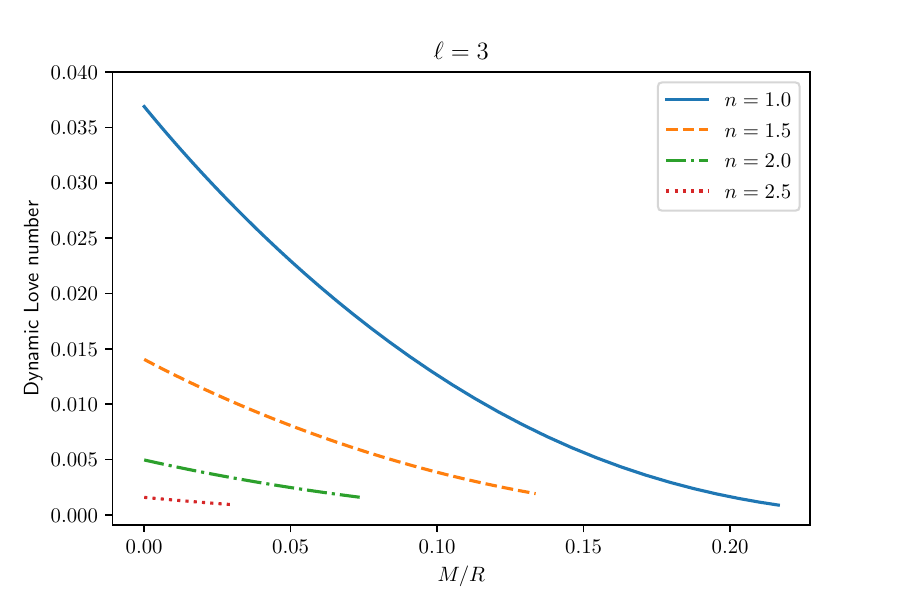}
\caption{Static and dynamic Love numbers for $\ell =3$, computed for a relativistic polytrope with an equation of state $p = K \rho^{1+1/n}$, where $n = \{1.0, 1.5, 2.0, 2.5\}$. Left panel: static Love number $k_\ell$. Right panel: dynamic Love number $\ddot{k}_\ell$. Each Love number is plotted as a function of the stellar compactness $M/R$. The curve ends at the configuration of maximum mass, beyond which the equilibrium sequence is dynamically unstable.}  
\label{fig:loveL3} 
\end{figure} 

\begin{figure}
\includegraphics[width=0.49\linewidth]{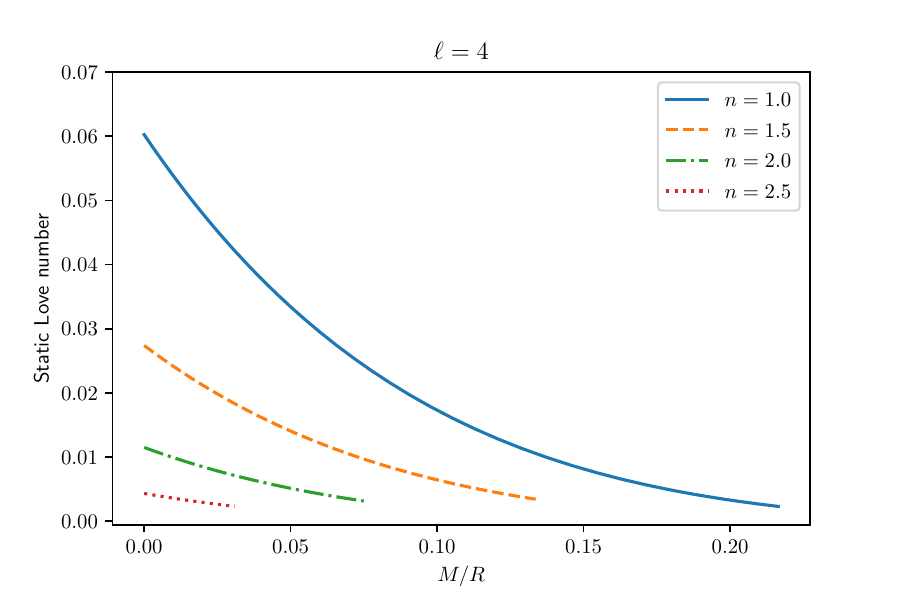}
\includegraphics[width=0.49\linewidth]{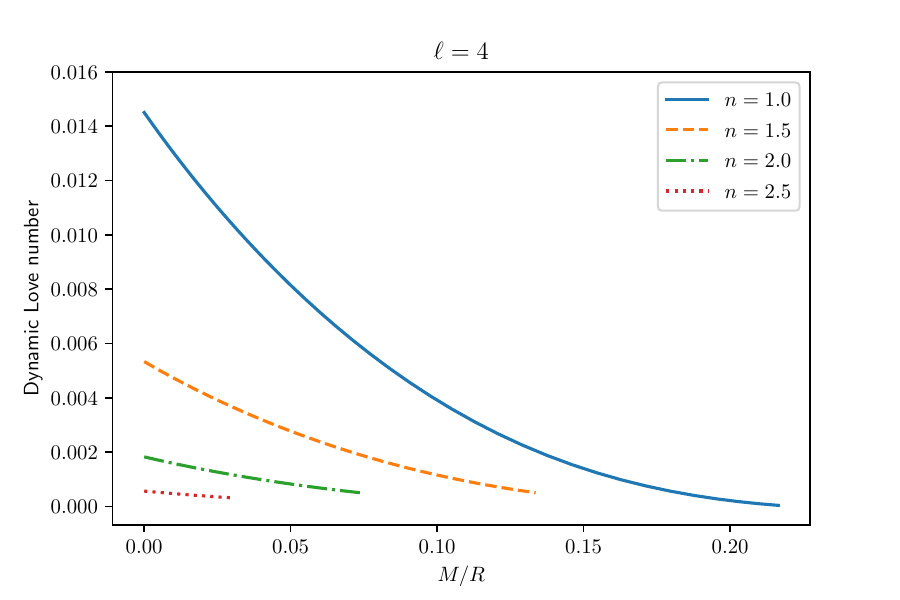}
\caption{Static and dynamic Love numbers for $\ell =4$, computed for a relativistic polytrope.}  
\label{fig:loveL4} 
\end{figure} 

\begin{figure}
\includegraphics[width=0.49\linewidth]{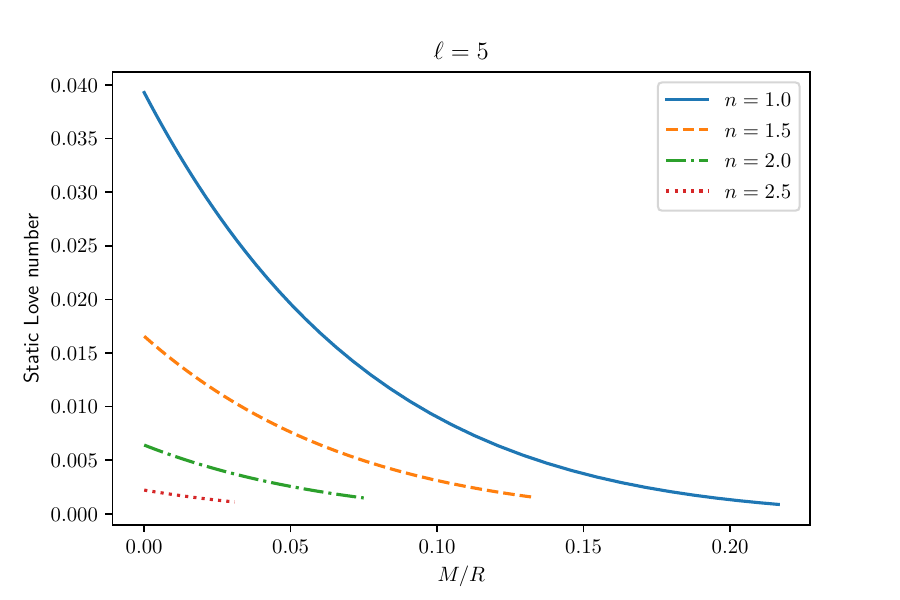}
\includegraphics[width=0.49\linewidth]{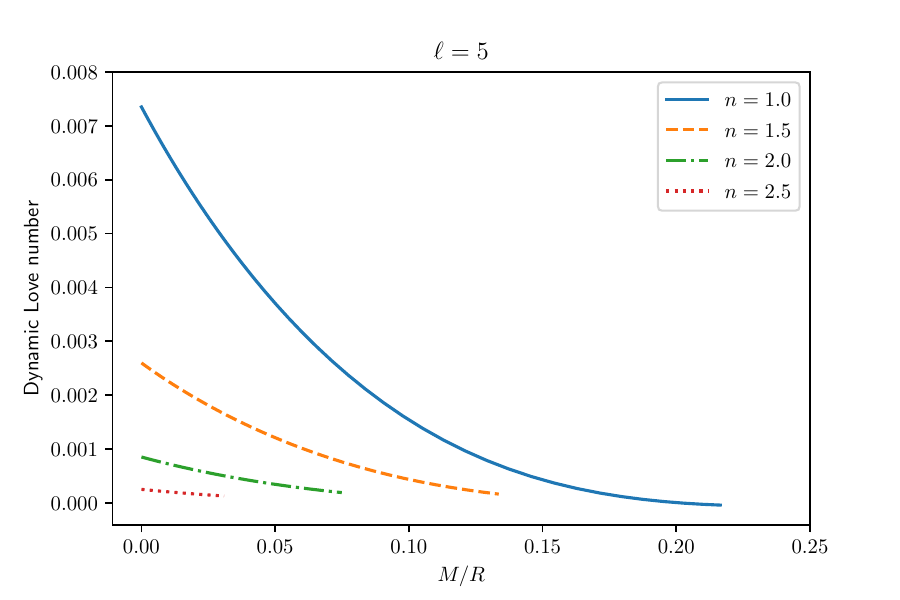}
\caption{Static and dynamic Love numbers for $\ell =5$, computed for a relativistic polytrope.}  
\label{fig:loveL5} 
\end{figure} 

Writing the perturbed metric as $g_{\alpha\beta} = g_{\alpha\beta}^0 + p_{\alpha\beta}$, where $g_{\alpha\beta}^0$ is the unperturbed metric and $p_{\alpha\beta}$ the perturbation, the matching conditions are
\begin{equation}
0 = \bigl[ g_{\alpha\beta}(r = R + \delta R) \bigr]
= \bigl[ g_{\alpha\beta}^0(r=R) \bigr] + \bigl[ \partial_r g_{\alpha\beta}^0(r=R) \bigr]\, \delta R
+\bigl[ p_{\alpha\beta}(r=R) \bigr],
\end{equation}
where $[g_{\alpha\beta}]$ is the difference between the exterior and interior metrics, and $r = R + \delta R$ marks the deformed stellar surface. Because the unperturbed metric is continuous and differentiable at $r = R$ \footnote{This is true when $\mu$ vanishes at $r=R$, which is the case for our polytropic equation of state.}, this simplifies to
\begin{equation}
\bigl[ p_{\alpha\beta}(r=R) \bigr] = 0.
\end{equation}
In words, the metric perturbation must be continuous at the unperturbed surface. 

We recall that a choice of normalization was made for the interior solution when we set $a_\ell(r=0) = 1$. As a consequence, the interior solution will differ from the exterior solution by an overall multiplicative factor, which we denote $N_\ell$. When we compare Eqs.~(\ref{metric_pert}) and (\ref{pert_inside}) and account for the fact that
$\exp(2\psi_s) = f_s = 1-2M/R$, we find that the matching conditions are
\begin{subequations}
\label{matching}
\begin{align}
N_\ell a_s &= e^s_{tt}, \\
N_\ell \bigl[ a_s + (R/r_1)^2 c_s \bigr] &= e^s, \\
N_\ell \ddot{a}_s &= (M/r_1)^2\, \ddot{e}^s_{tt}, \\
N_\ell \bigl[ \ddot{a}_s + (R/r_1)^2 \ddot{c}^s \bigr] &= (M/r_1)^2\, \ddot{e}^s,
\end{align}
\end{subequations}
as well as $N_\ell \dot{b}_s = \dot{e}^s_{tr}$. We use the notation $a_s := a_\ell(r=R)$, $e^s_{tt} := e_{tt}(r=R)$, and so on for the remaining radial functions. 

\begin{table} 
\caption{\label{tab:k} Static Love numbers for a polytrope with an equation of state $p = K \rho^{1+1/n}$. For each value of $n$ we provide five values between $b = 10^{-5}$ and $b = b_{\rm max}$, where $b := p(r=0)/\rho(r=0)$ parametrizes the equilibrium sequence, and where $b_{\rm max}$ denotes the value at the configuration of maximum mass. The entries are equally spaced in $\log_{10} b$. A comparison between two independent sets of computations reveals agreement for all 6 significant digits.} 
\begin{ruledtabular}
\begin{tabular}{ccccccc}
$n$ & $b$ & $M/R$ & $k_2$ & $k_3$ & $k_4$ & $k_5$ \\ 
\hline 
1.0 & 1.00000\e{-5}  &  1.99989\e{-5}  & 2.59872\e{-1} & 1.06435\e{-1} & 6.02278\e{-2} & 3.92821\e{-2} \\ 
& 1.34661\e{-4}  &  2.69119\e{-4}  & 2.59419\e{-1} & 1.06193\e{-1} & 6.00598\e{-2} & 3.91523\e{-2} \\
& 1.81337\e{-3}  &  3.59016\e{-3}  & 2.53428\e{-1} & 1.03011\e{-1} & 5.78571\e{-2} & 3.74579\e{-2} \\
& 2.44191\e{-2}  &  4.28971\e{-2}  & 1.89453\e{-1} & 7.06771\e{-2} & 3.64705\e{-2} & 2.17086\e{-2} \\
& 3.28830\e{-1}  &  2.16430\e{-1}  & 2.85515\e{-2} & 6.99089\e{-3} & 2.31968\e{-3} & 8.84569\e{-4} \\
  & & & & & & \\
1.5 & 1.00000\e{-5}  & 1.85691\e{-5}  & 1.43257\e{-1} & 5.28383\e{-2} & 2.73867\e{-2} & 1.65643\e{-2} \\
& 1.10349\e{-4}  & 2.04774\e{-4}  & 1.43038\e{-1} & 5.27362\e{-2} & 2.73229\e{-2} & 1.65192\e{-2} \\ 
& 1.21768\e{-3}  & 2.24345\e{-3}  & 1.40650\e{-1} & 5.16277\e{-2} & 2.66327\e{-2} & 1.60330\e{-2} \\ 
& 1.34370\e{-2}  & 2.29261\e{-2}  & 1.17851\e{-1} & 4.13422\e{-2} & 2.03934\e{-2} & 1.17457\e{-2} \\ 
& 1.48275\e{-1}  & 1.33622\e{-1}  & 3.32914\e{-2} & 9.05800\e{-3} & 3.43752\e{-3} & 1.52301\e{-3} \\
  & & & & & & \\
2.0 & 1.00000\e{-5}  & 1.66157\e{-5}  & 7.39258\e{-2} & 2.43890\e{-2} & 1.15050\e{-2} & 6.41818\e{-3} \\ 
& 9.21652\e{-5}  & 1.53042\e{-4}  & 7.38227\e{-2} & 2.43476\e{-2} & 1.14820\e{-2} & 6.40350\e{-3} \\ 
& 8.49442\e{-4}  & 1.40241\e{-3}  & 7.28823\e{-2} & 2.39709\e{-2} & 1.12735\e{-2} & 6.27031\e{-3} \\ 
& 7.82890\e{-3}  & 1.22690\e{-2}  & 6.49915\e{-2} & 2.08640\e{-2} & 9.57982\e{-3} & 5.20403\e{-3} \\ 
& 7.21552\e{-2}  & 7.49332\e{-2}  & 2.85166\e{-2} & 7.95156\e{-3} & 3.15909\e{-3} & 1.48641\e{-3} \\
  & & & & & & \\
2.5 & 1.00000\e{-5}  & 1.42934\e{-5}  & 3.48455\e{-2} & 1.01897\e{-2} & 4.34038\e{-3} & 2.21950\e{-3} \\ 
& 7.23073\e{-5}  & 1.03293\e{-4}  & 3.48030\e{-2} & 1.01752\e{-2} & 4.33332\e{-3} & 2.21545\e{-3} \\
& 5.22834\e{-4}  & 7.43820\e{-4}  & 3.44980\e{-2} & 1.00712\e{-2} & 4.28278\e{-3} & 2.18650\e{-3} \\ 
& 3.78047\e{-3}  & 5.22233\e{-3}  & 3.24007\e{-2} & 9.36235\e{-3} & 3.94083\e{-3} & 1.99191\e{-3} \\ 
& 2.73356\e{-2}  & 3.08993\e{-2}  & 2.15146\e{-2} & 5.86403\e{-3} & 2.32536\e{-3} & 1.10848\e{-3} 
\end{tabular} 
\end{ruledtabular} 
\end{table} 

\begin{table} 
\caption{\label{tab:kddot} Dynamic Love numbers for a polytrope with an equation of state $p = K \rho^{1+1/n}$. A comparison between two independent sets of computations reveals that $\ddot{k}_2$ is accurate through 6 significant digits, that $\ddot{k}_3$ is accurate through 5 significant digits, and that $\ddot{k}_4$ and $\ddot{k}_5$ are accurate through 4 digits. For $n=1$ and $\ell = 5$ we find that $\ddot{k}_\ell$ appears to change sign just as $M/R$ approaches its final value. We do not believe that this behavior is physical, but more likely a numerical artefact. We have not, however, been able to pin point the source of the numerical error.} 
\begin{ruledtabular}
\begin{tabular}{ccccccc}
$n$ & $b$ & $M/R$ & $\ddot{k}_2$ & $\ddot{k}_3$ & $\ddot{k}_4$ & $\ddot{k}_5$ \\ 
\hline 
1.0 & 1.00000\e{-5}  &  1.99989\e{-5}  & 1.72580\e{-1}  & 3.68781\e{-2} & 1.45070\e{-2} & 7.35112\e{-3} \\ 
& 1.34661\e{-4}  &  2.69119\e{-4}  & 1.72266\e{-1}  & 3.67978\e{-2} & 1.44686\e{-2} & 7.32804\e{-3} \\
& 1.81337\e{-3}  &  3.59016\e{-3}  & 1.68112\e{-1}  & 3.57385\e{-2} & 1.39644\e{-2} & 7.02600\e{-3} \\
& 2.44191\e{-2}  &  4.28971\e{-2}  & 1.23566\e{-1}  & 2.46835\e{-2} & 8.90707\e{-3} & 4.12171\e{-3} \\
& 3.28830\e{-1}  &  2.16430\e{-1}  & 1.13698\e{-2}  & 8.97871\e{-4} & 3.79990\e{-5} & $-$4.00445\e{-5} \\
  & & & & & & \\
1.5 & 1.00000\e{-5}  & 1.85691\e{-5}  & 6.74485\e{-2}  &  1.40504\e{-2}  & 5.33396\e{-3}  & 2.59909\e{-3} \\
& 1.10349\e{-4}  & 2.04774\e{-4}  & 6.73393\e{-2}  &  1.40243\e{-2}  & 5.32220\e{-3}  & 2.59239\e{-3} \\ 
& 1.21768\e{-3}  & 2.24345\e{-3}  & 6.61507\e{-2}  &  1.37406\e{-2}  & 5.19472\e{-3}  & 2.51995\e{-3} \\ 
& 1.34370\e{-2}  & 2.29261\e{-2} & 5.48014\e{-2}  &  1.10729\e{-2}  & 4.02294\e{-3}  & 1.86992\e{-3} \\ 
& 1.48275\e{-1}  & 1.33622\e{-1}  & 1.22637\e{-2}  &  1.93090\e{-3}  & 5.03042\e{-4}  & 1.63785\e{-4} \\
  & & & & & & \\
2.0 & 1.00000\e{-5}  & 1.66157\e{-5}  & 2.41628\e{-2}  & 4.97197\e{-3}  & 1.82433\e{-3}  & 8.52832\e{-4} \\ 
& 9.21652\e{-5}  & 1.53042\e{-4}  & 2.41264\e{-2}  & 4.96380\e{-3}  & 1.82088\e{-3}  & 8.50982\e{-4} \\ 
& 8.49442\e{-4}  & 1.40241\e{-3}  & 2.37952\e{-2}  & 4.88944\e{-3}  & 1.78948\e{-3}  & 8.34185\e{-4} \\ 
& 7.82890\e{-3}  & 1.22690\e{-2}  & 2.10217\e{-2}  & 4.27203\e{-3}  & 1.53222\e{-3}  & 6.98461\e{-4} \\ 
& 7.21552\e{-2}  & 7.49332\e{-2}  & 8.23899\e{-3}  & 1.55958\e{-3}  & 4.88879\e{-4}  & 1.92052\e{-4} \\
  & & & & & & \\
2.5 & 1.00000\e{-5}  & 1.42934\e{-5}  & 7.74388\e{-3}  & 1.59489\e{-3}  & 5.64580\e{-4}  & 2.51983\e{-4} \\ 
& 7.23073\e{-5}  & 1.03293\e{-4}  & 7.73352\e{-3}  & 1.59268\e{-3}  & 5.63705\e{-4}  & 2.51546\e{-4} \\
& 5.22834\e{-4}  & 7.43820\e{-4} & 7.65915\e{-3}  & 1.57679\e{-3}  & 5.57433\e{-4}  & 2.48416\e{-4} \\ 
& 3.78047\e{-3}  & 5.22233\e{-3}  & 7.14925\e{-3}  & 1.46820\e{-3}  & 5.14838\e{-4}  & 2.27297\e{-4} \\ 
& 2.73356\e{-2}  & 3.08993\e{-2}  & 4.53405\e{-3}  & 9.21156\e{-4}  & 3.07954\e{-4}  & 1.28630\e{-4} 
\end{tabular} 
\end{ruledtabular} 
\end{table} 

Making use of Eq.~(\ref{static_functions}), we have that the first two equations in Eq.~(\ref{matching}) take the explicit form
\begin{subequations}
\label{match_static}
\begin{align}
0 &= 2 B_\ell\, k_\ell - a_\ell\, N_\ell + A_\ell, \\
0 &= 2 D_\ell\, k_\ell - \bigl[ a_\ell + (R/r_1)^2 c_\ell \bigr]\, N_\ell + C_\ell,
\end{align}
\end{subequations}
in which all radial functions are evaluated at $r=R$. We have two equations for the two unknowns $k_\ell$ and $N_\ell$. It is useful to note that $(R/r_1)^2 = b \zeta_s$.  

With Eq.~(\ref{dynamic_functions}) we have that the last two equations in Eq.~(\ref{matching}) can be expressed as
\begin{subequations}
\label{match_dynamic} 
\begin{align}
0 &= A_\ell\, \ddot{t}_\ell - 2B_\ell\, \ddot{k}_\ell - (M/R) \Bigl\{ N_\ell (r_1/R)^2 \ddot{a}_\ell
- (M/R)^2 \gotha_\ell - k_\ell (R/M)^{2\ell-1} \gothb_\ell \Bigr\}, \\
0 &= C_\ell\, \ddot{t}_\ell - 2D_\ell\, \ddot{k}_\ell - (M/R) \Bigl\{
  N_\ell \bigl[ (r_1/R)^2 \ddot{a}_\ell + \ddot{c}_\ell \bigr] 
  - (M/R)^2 \gothc_\ell - k_\ell (R/M)^{2\ell-1} \gothd_\ell \Bigr\},
\end{align}
\end{subequations}
in which the radial functions are again evaluated at $r=R$. Here we have two equations for the two unknowns $\ddot{k}_\ell$ and $\ddot{t}_\ell$.

The remaining matching condition, $N_\ell \dot{b}_s = \dot{e}^s_{tr}$, brings no new information. It is nevertheless useful, because it can be turned into a test of the numerics. We have verified that the equation holds up to the degree of accuracy expected of our computations.  

As was explained in the paragraph following Eq.~(\ref{moment_redefinition2}), our ultimate goal is to construct an exterior solution with $\ddot{t}_\ell := (M/R)^3 \ddot{T}_\ell$ set to zero. To achieve this we exercise the freedom to redefine the tidal moments according to Eq.~(\ref{moment_redefinition1}), which produces the changes described by Eq.~(\ref{moment_redefinition2}). Selecting $\ddot{\lambda}_\ell$ so that the new $\ddot{t}_\ell$ vanishes, we find that the new dynamic Love number is given by 
\begin{equation}
\ddot{k}^{\rm new}_\ell = \ddot{k}^{\rm old}_\ell + k_\ell \ddot{t}^{\rm old}_\ell,
\label{old_new} 
\end{equation}
with the right-hand side featuring the old values returned by the computation. At this stage the freedom to redefine the tidal moments is exhausted, and the new and final Love number is invariant. 

The results of our computations are presented in Figs.~\ref{fig:loveL2}, \ref{fig:loveL3}, \ref{fig:loveL4}, \ref{fig:loveL5}, as well as in Tables~\ref{tab:k} and \ref{tab:kddot}. We observe that the values for $M/R \ll 1$ agree very well with the Newtonian results displayed in Table~\ref{tab:love}. We notice also that the Love numbers decrease with increasing $M/R$, reaching a minimum when the equilibrium sequence comes to an end at the configuration of maximum mass.

The computation of the static and dynamic Love numbers requires an accurate evaluation of the radial functions that appear in Eqs.~(\ref{match_static}) and (\ref{match_dynamic}), from $A_\ell$ through $\gothd_\ell$. Some words of advice regarding this task are offerred in Appendix~\ref{sec:numerical}. 

\begin{acknowledgments} 
This work was supported by the Natural Sciences and Engineering Research Council of Canada.  
\end{acknowledgments} 

\appendix

\section{Numerical techniques}
\label{sec:numerical}

We provide brief descriptions of our numerical methods in this Appendix. We begin in Sec.~\ref{subsec:collocation} with a presentation of the collocation methods that were used in the integration of the structure and perturbation equations in Secs.~\ref{sec:newtonian} and \ref{sec:GR}. This is followed by a description of our finite-difference methods in Sec~\ref{subsec:finite}. In Sec.~\ref{subsec:evaluation} we explain our use of continued-fraction representations of some radial functions to provide a numerically reliable evaluation.  

\subsection{Collocation}
\label{subsec:collocation} 

We use a collocation method for the Newtonian computations of Sec.~\ref{sec:newtonian} and the relativistic computations of Sec.~\ref{sec:GR}. This is based on an expansion of all variables in Chebyshev polynomials \cite{boyd:01}, with the differential equations giving rise to a system of algebraic equations for the coefficients. Here, as an illustration, we describe the method in the simplified context of a static tidal perturbation in Newtonian theory. The method extends easily to the dynamic problem, and to the relativistic formulation. 

The system of equations is provided by Eqs.~(\ref{e_eqns_dyn}) and (\ref{v_eqns_dyn}), 
\begin{subequations}
\begin{align}
0 &= E^e := r \frac{d e}{dr} - v, \\ 
0 &= E^v := r \frac{d v}{dr} + (2\ell+1) v + 6 n \zeta \vartheta^{n-1}\, e,
\end{align}
\end{subequations}
with $e := e^0_\ell$ and $v := v^0_\ell$, and with 
\begin{equation}
r \frac{d}{dr} = -2\zeta \chi \frac{d}{d\vartheta}. 
\end{equation}
The system comes with boundary conditions $v(\vartheta = 1) = 0$ and $e(\vartheta = 0) = 1$. We assume that the structure functions $\zeta(\vartheta)$ and $\chi(\vartheta)$ were previously computed --- they must also be obtained numerically --- for a polytrope with index $n$. We see that the equation $E^v = 0$ is singular at $\vartheta = 0$ when $n < 1$; we exclude such cases from our considerations.

The Chebyshev polynomials $T_p(x) := \cos(pt)$, with $p = 0, 1, 2, \cdots$ and $x = \cos t$, are defined in the interval $-1 \leq x \leq 1$. We therefore rescale $\vartheta$ according to
\begin{equation}
\vartheta = \frac{1}{2} (x + 1).
\end{equation}
The center at $\vartheta = 1$ is mapped to $x = 1$ and $t = 0$, while the surface at $\vartheta = 0$ is mapped to $x = -1$ and $t = \pi$. The dependent variables are expanded as 
\begin{equation}
e = \sum_{p=0}^{N-1} e_p\, \cos(pt), \qquad
v = \sum_{p=0}^{N-1} v_p\, \cos(pt),
\end{equation}
where $e_p$, $v_p$ are constants, and $N$ is the total number of terms kept in the expansions. The derivatives with respect to $x$ are then
\begin{equation}
\frac{de}{dx} = \frac{1}{\sin t} \sum_{p=0}^{N-1} p e_p\, \sin(pt), \qquad
\frac{dv}{dx} = \frac{1}{\sin t} \sum_{p=0}^{N-1} p v_p\, \sin(pt),
\end{equation}
and the radial derivative is now written as $r (d/dr) = -4\zeta \chi (d/dx)$.

To obtain the coefficients $e_p$, $v_p$ we turn the differential equations and boundary conditions into a set of $2N$ algebraic equations. We write
\begin{equation}
E^e_k := E^e(t = t_k) = 0, \qquad 
E^v_k := E^v(t = t_k) = 0,
\end{equation}
where
\begin{equation}
t_k := \frac{(k-\tfrac{1}{2}) \pi}{N-1}, \qquad k = 1, 2, \cdots, N-1
\end{equation}
are the collocation points, given by all the zeros of $T_N(x)$ --- this is the Gauss-Chebyshev grid. Because there are $N-1$ collocation points, the algebraic system includes $2N - 2$ equations so far. The remaining two equations are supplied by the boundary conditions,
\begin{equation}
v(t = 0) = 0, \qquad
e(t = \pi) = 1.
\end{equation}
We have $2N$ algebraic equations for the the $2N$ unknowns, and the solution can be found with standard techniques from linear algebra.

The collocation method can also be exploited to compute the structure functions $\zeta(\vartheta)$ and $\chi(\vartheta)$. Here the situation is more complicated, because the structure equations are nonlinear, while the collocation method, with its reliance on linear algebra, works best with linear problems. An effective approach is to linearize the equations about a guessed approximation to the solution, and to perform iterations to improve the approximation. 

\subsection{Finite differences}
\label{subsec:finite} 

We also employ finite-difference methods to integrate the perturbation equations in their relativistic formulation. We provide some details of implementation here.

The integration of Eqs.~(\ref{0time}) for $a_\ell$ and $c_\ell$ comes with no particular challenge, and the equations can be integrated straightforwardly from (almost) $r=0$ to $r=R$. As we explained in the main text, we actually use $\vartheta$, a substitute for the mass density $\rho$, as the independent variable, which ranges from $\vartheta = 1$ at $r=0$ to $\vartheta = 0$ at $r = R$. The point $r = 0$ must be excluded from the computation because the differential equations are singular there. To account for this we express all variables as Taylor series about $r=0$, and substitute these within the differential equations to obtain starting values at $\vartheta = 1 - \varepsilon$, where $\varepsilon$ is chosen to be numerically small. 

The integration of Eqs.~(\ref{1time1}) for $\dot{u}_\ell$ and $\dot{v}_\ell$ is more challenging, because the system is singular at both $r=0$ and $r=R$, and we require a boundary condition at both $r = 0$ and $r = R$; refer to Eqs.~(\ref{BD_centre_b}) and (\ref{BD_surface}). To handle this we perform two integrations, the first from (almost) $r= 0$ up to a middle point $r = r^\sharp$ corresponding to $\vartheta = 1/2$, and the second from (almost) $r=R$ down to the middle point. In this case also we perform Taylor expansions to obtain starting values at $\vartheta = 1 - \varepsilon$ and $\vartheta = \varepsilon$. \footnote{Strictly speaking, an expansion in powers of $\vartheta$ near the surface must also include fractional powers when the polytropic index $n$ is not an integer. When $n \geq 1$, as we assume here, this leads to complications at higher powers of $\vartheta$ than are required in our computations, and the issue can be ignored.} 

For the inner integration the correct value of $\dot{u}_0 := \dot{u}_\ell(r=0)$ is unknown, and to accommodate our ignorance we construct a general solution to the differential equations for an arbitrary value of this quantity. Writing $\bm{u} := (\dot{u}_\ell, \dot{v}_\ell)$ and exploiting the linearity of the equations, we write
\begin{equation}
\bm{u}_{\rm inner} = \frac{1}{2} \dot{u}_0 \bigl( \bm{u}_{\rm inner}^{\rm I} - \bm{u}_{\rm inner}^{\rm II} \bigr)
+ \frac{1}{2} \bigl( \bm{u}_{\rm inner}^{\rm I} + \bm{u}_{\rm inner}^{\rm II} \bigr),
\end{equation}
where the basis of independent solutions is defined by
\begin{equation}
\bm{u}_{\rm inner}^{\rm I} := \bm{u}(\dot{u}_0 = 1), \qquad
\bm{u}_{\rm inner}^{\rm II} := \bm{u}(\dot{u}_0 = -1).
\end{equation}
Two inner integrations return $\bm{u}^{\rm I}_{\rm inner}$ and $\bm{u}^{\rm II}_{\rm inner}$, and $\bm{u}_{\rm inner}$ is known up to the value of $\dot{u}_0$.  

For the outer integration it is the correct value of $\dot{v}_s := \dot{v}_\ell(r=R)$ that is unknown, and we construct a general solution by writing
\begin{equation}
\bm{u}_{\rm outer} = \frac{1}{2} \dot{v}_s \bigl( \bm{u}_{\rm outer}^{\rm I} - \bm{u}_{\rm outer}^{\rm II} \bigr)
+ \frac{1}{2} \bigl( \bm{u}_{\rm outer}^{\rm I} + \bm{u}_{\rm outer}^{\rm II} \bigr), 
\end{equation}
where
\begin{equation}
\bm{u}_{\rm outer}^{\rm I} := \bm{u}(\dot{v}_s = 1), \qquad
\bm{u}_{\rm outer}^{\rm II} := \bm{u}(\dot{v}_s = -1). 
\end{equation}
Two outer integrations give us the basis functions, and $\bm{u}_{\rm outer}$ is known up to the value of $\dot{v}_s$.

Continuity at $r = r^\sharp$,
\begin{equation}
\bm{u}_{\rm inner}(r=r^\sharp) = \bm{u}_{\rm outer}(r=r^\sharp),
\end{equation}
provides us with two equations for the two unknowns $\dot{u}_0$ and $\dot{v}_s$. Solving for these gives us the correct global solution for $\dot{u}_\ell$ and $\dot{v}_\ell$.

The integration of Eqs.~(\ref{2time}) for $\ddot{a}_\ell$ and $\ddot{c}_\ell$ is straightforward in principle, because the equations are regular at $r = R$, and $\ddot{a}_\ell(r=0) = 1$ is the only required boundary condition; see Eq.~(\ref{BD_centre_c}). The equations, however, must be integrated simultaneously with Eqs.~(\ref{0time}) and (\ref{1time1}), and the latter set is singular at $r=R$. It is therefore preferable to adopt the same practice as with Eqs.~(\ref{1time1}), and to carry out inner and outer integrations.

For the inner integration the situation is simple, because (as was just stated) the choice $\ddot{a}_\ell(r=0) = 1$ specifies a unique solution. For the outer solution we have the freedom to specify both $\ddot{a}_s := \ddot{a}_\ell(r=R)$ and $\ddot{c}_s := \ddot{c}_\ell(r=R)$, and a general solution to the differential equations will be linear in both quantities. It is easy to show that the general solution can be expressed as
\begin{equation}
\bm{a}_{\rm outer} = \ddot{a}_s \bigl( \bm{a}_{\rm outer}^{\rm III} - \bm{a}_{\rm outer}^{\rm II} \bigr)
+ \ddot{c}_s \bigl( \bm{a}_{\rm outer}^{\rm III}- \bm{a}_{\rm outer}^{\rm I} \bigr)
+ \bm{a}_{\rm outer}^{\rm I} + \bm{a}_{\rm outer}^{\rm II} - \bm{a}_{\rm outer}^{\rm III},
\end{equation}
where the basis of independent solutions is defined by
\begin{equation}
\bm{a}_{\rm outer}^{\rm I} := \bm{a}\bigl( \ddot{a}_s=1, \ddot{c}_s=0 \bigr), \qquad
\bm{a}_{\rm outer}^{\rm II} := \bm{a}\bigl( \ddot{a}_s=0, \ddot{c}_s=1 \bigr), \qquad
\bm{a}_{\rm outer}^{\rm III} := \bm{a}\bigl( \ddot{a}_s=1, \ddot{c}_s=1 \bigr).
\end{equation}
Three outer integrations return the basis functions, and $\bm{a}_{\rm outer}$ is known up to the values of $\ddot{a}_s$ and $\ddot{c}_s$.

Continuity at $r = r^\sharp$,
\begin{equation}
\bm{a}_{\rm inner}(r = r^\sharp) = \bm{a}_{\rm outer}(r = r^\sharp),
\end{equation}
gives us two equations for the two unknowns $\ddot{a}_s$ and $\ddot{c}_s$, and we arrive at the correct global solution for $\ddot{a}_\ell$ and $\ddot{c}_\ell$.

\subsection{Evaluation of radial functions}
\label{subsec:evaluation} 

The computation of the relativistic Love numbers in Sec.~\ref{sec:GR} requires the evaluation of a large number of functions of $M/r$ at the stellar surface $r = R$. We require a computational method that allows us to probe many orders of magnitude for $M/R$, going from very low values in the Newtonian limit, to approximately $M/R = 0.3$ for the most compact stellar models.

The functions $A_\ell$, $B_\ell$, $C_\ell$, and $D_\ell$ are defined in terms of hypergeometric functions in Eq.~(5.8) of Ref.~\cite{poisson:21a}. The last argument of these functions is $2M/R$, and we observe that the hypergeometric series converges rapidly for all sampled values of $M/R$. The definition, therefore, provides an efficient and reliable means of computation. For specific values of $\ell$, which we take to be in the set $\ell = \{2,3,4,5\}$, the hypergeometric functions can also be expressed in terms of simple functions. We find that $A_\ell$ and $C_\ell$ are terminating polynomials, while $B_\ell$ and $D_\ell$ involve polynomials and the function $\ln(1-2M/R)$. If we begin with these explicit expressions, we find that values for $A_\ell$ and $C_\ell$ can be obtained reliably by direct evaluation. The same is true for $B_\ell$ and $D_\ell$ when $M/R$ is not too small. But when $M/R$ is small, we observe that the numerical accuracy quickly degrades because of near cancellations between the logarithmic and polynomial terms; many, and sometimes all, significant digits are lost in the operation. In such cases we require an alternative method to compute $B_\ell$ and $D_\ell$.

We may of course return to the representation in terms of hypergeometric functions. An alternative method turns out to be equally efficient and reliable, and can be adopted for the remaining radial functions. We proceed as follows. Consider, for example, the specific case of $B_2$. We write down its explicit expression in terms of polynomials and logarithm, and with a symbolic manipulation software (we use Maple), we carry out a Taylor expansion in powers of $M/R$, up through order $(M/R)^{32}$ (because we can). This representation is extremely accurate for small values of $M/R$, but it is entirely useless when $M/R$ is comparable to $0.3$. To repair this we re-express the Taylor expansion as an equivalent continued fraction. As an illustration, we would write
\begin{equation}
\ln(1+x) = x - \frac{1}{2} x^2 + \frac{1}{3} x^3 + O(x^4)
= \cfrac{x}{1 + \cfrac{x}{2 + \cfrac{x}{3 + \cdots}}} + O(x^4), 
\end{equation}
except that our expansions are actually much longer, with an error term of $O(x^{33})$. The representation of $B_2$ as a continued fraction turns out to be machine-precision accurate (at least 15 significant digits) in the specified range of $M/R$. This is true for all the other instances of $B_\ell$, and for $D_\ell$ as well.

The same considerations apply to $\dot{e}_{tr}$. This is defined in terms of hypergeometric functions in Eq.~(5.47) of Ref.~\cite{poisson:21a}, and listed explicitly for $\ell = \{2,3,4,5\}$ in Appendix E of this reference. Here also one is given the choice between hypergeometric and continued-fraction representations. Both are accurate to machine precision in the relevant interval of $M/R$.

The continued-fraction representation is the only viable option in the case of the remaining radial functions, $\gotha_\ell$, $\gothb_\ell$,  $\gothc_\ell$, and $\gothd_\ell$, which are listed explicitly for $\ell = \{2,3,4,5\}$ in Appendix F of Ref.~\cite{poisson:21a}. These functions are defined in terms of polynomials, logarithms, and polylogarithms, and they are complicated. A direct evaluation from the explicit expressions reveals a severe loss of numerical accuracy when $M/R$ is small. We therefore proceed as before, with an expansion through $(M/R)^{32}$ 
and a conversion to an equivalent continued fraction. The manipulations are complicated by the fact that the expansion of each radial function contains two pieces, one a straight polynomial in $M/R$, the other a polynomial in $M/R$ multiplying $\ln(M/R)$. To account for this we convert each polynomial into a continued fraction, and express the function in the form
\begin{equation}
(\mbox{continued fraction in $M/R$}) + (\mbox{continued fraction in $M/R$}) \ln(M/R).
\end{equation}
With this representation we again achieve machine precision in the interval $0 < M/R < 0.3$.

\bibliography{love_dyn}
\end{document}